\documentclass[reprint, showkeys,%
longbibliography,onecolumn,notitlepage,
amsmath,amssymb,
 aps,
 prf,
]{revtex4-1}
\usepackage{amsmath}
\usepackage{graphicx}
\usepackage{dcolumn}
\usepackage{bm}
\usepackage{xcolor}

\begin{document}
\title{On Lagrangian Intermittency from 4-D Particle Tracking Velocimetry measurements of a turbulent von K\'{a}rm\'{a}n flow}

\author{Valentina Valori}
\affiliation{SPEC, CEA, CNRS, Universit\'{e} Paris-Saclay, CEA Saclay, Gif-sur-Yvette, France}%
\email{valentina.valori@tu-ilmenau.de}
\author{Paul Debue}%
\affiliation{SPEC, CEA, CNRS, Universit\'{e} Paris-Saclay, CEA Saclay, Gif-sur-Yvette, France}%
\author{Adam Cheminet}
\affiliation{SPEC, CEA, CNRS, Universit\'{e} Paris-Saclay, CEA Saclay, Gif-sur-Yvette, France}%
\author{Tarek Chaabo}
\affiliation{Univ. Lille, CNRS, ONERA, Arts et Metiers Institute of Technology, Centrale Lille, UMR 9014-LMFL-Laboratoire de M\'ecanique des Fluides de Lille-Kamp\'e de F\'eriet, F-59000, Lille, France}
\author{Ya\c{s}ar Ostovan}
\affiliation{Univ. Lille, CNRS, ONERA, Arts et Metiers Institute of Technology, Centrale Lille, UMR 9014-LMFL-Laboratoire de M\'ecanique des Fluides de Lille-Kamp\'e de F\'eriet, F-59000, Lille, France}
\author{Christophe Cuvier}
\affiliation{Univ. Lille, CNRS, ONERA, Arts et Metiers Institute of Technology, Centrale Lille, UMR 9014-LMFL-Laboratoire de M\'ecanique des Fluides de Lille-Kamp\'e de F\'eriet, F-59000, Lille, France}
\author{Jean-Marc Foucaut}
\affiliation{Univ. Lille, CNRS, ONERA, Arts et Metiers Institute of Technology, Centrale Lille, UMR 9014-LMFL-Laboratoire de M\'ecanique des Fluides de Lille-Kamp\'e de F\'eriet, F-59000, Lille, France}
\author{Jean-Philippe Laval}
\affiliation{Univ. Lille, CNRS, ONERA, Arts et Metiers Institute of Technology, Centrale Lille, UMR 9014-LMFL-Laboratoire de M\'ecanique des Fluides de Lille-Kamp\'e de F\'eriet, F-59000, Lille, France}
\author{C\'{e}cile Wiertel}%
\affiliation{SPEC, CEA, CNRS, Universit\'{e} Paris-Saclay, CEA Saclay, Gif-sur-Yvette, France}%
\author{Vincent Padilla}%
\affiliation{SPEC, CEA, CNRS, Universit\'{e} Paris-Saclay, CEA Saclay, Gif-sur-Yvette, France}%
\author{Fran\c{c}ois Daviaud}%
\affiliation{SPEC, CEA, CNRS, Universit\'{e} Paris-Saclay, CEA Saclay, Gif-sur-Yvette, France}%
\author{B\'{e}reng\`{e}re Dubrulle}%
\affiliation{CNRS, SPEC, CEA, Universit\'{e} Paris-Saclay, CEA Saclay, Gif-sur-Yvette, France}%

\date{\today}

\begin{abstract}
We investigate the ability of 4D Particle Tracking Velocimetry measurements at high particle density to explore intermittency and irreversibility in a turbulent swirling flow at various Reynolds numbers. For this, we devise suitable tools to remove the experimental noise, and compute   the statistics of both Lagrangian velocity increments and wavelet coefficients of the Lagrangian power (the time derivative of the kinetic energy along a trajectory). We show that the signature of noise is strongest on short trajectories, and results in deviations from the regularity condition at small time scales.
Considering only long trajectories to get rid of such effect, we obtain scaling regimes that are compatible with a reduced intermittency, meaning that long trajectories are also associated with areas of larger regularity. The scaling laws, both in time and Reynolds number, can be described by the multifractal model, with a log-normal spectrum and an intermittency parameter that is three times smaller than in the Eulerian case, where all the areas of the flow are taken into account.
\end{abstract}

\maketitle

\section{\label{introduction}Introduction}
Intermittency is an important property of turbulent flow. It is manifested by highly inhomogeneous distribution of enstrophy, that results in highly fluctuating instantaneous local energy dissipation, inspiring the Kolmogorov-Obukhov (KO62)  explanation for observed deviations of self-similarity of Eulerian velocity structure functions \cite{K62}. This means that the scale symmetry breaking of the velocity structure functions is linked with the time-reversal symmetry breaking of the velocity field, or, that Eulerian intermittency is connected to irreversibility.
Manifestations of intermittency in the Lagrangian framework have also be sought at the level of individual fluid particles, by analysis of trajectories of tracers passively advected by the flow. Structure functions of the velocity along the trajectories $v(x(t))$  were also observed to deviate from a self-similar behaviour \cite{Mordant,Biferale}.
Chevillard et al \cite{Chevillard} showed using multi-fractal theory and Taylor hypothesis that such deviations can be simply mapped to the corresponding deviations in the Eulerian velocity field. By analogy with the Eulerian case, one can also investigate the connection between such intermittency and irreversibility \cite{Jucha}. A noticeable asymmetry property of Lagrangian trajectories was discovered by looking at the behaviour of the local kinetic energy along the trajectory $E=v(x(t))^2/2$. It is characterised by "flight-crash" events, a slow growth of $E(t)$ followed by a rapid decay \cite{pumirbodenschatz}. This motivated the analysis of the Lagrangian power, i.e. the instantaneous kinetic energy dissipated along the trajectory $P(t)=v\cdot dv/dt$. Such quantity was observed to be asymmetric with respect to time-reversal, resulting in a non-zero $<P^3>$, that becomes increasingly negative as the Reynolds number is increased: $<P^3>\sim -Re_\lambda^{2}$ \cite{Xu,Pumir}.  More generally, its moments of order $q$  display a power-law dependence on the Reynolds number  $<P^q>\sim -Re_\lambda^{\alpha(q)}$, that deviates from a self-similar prediction $\alpha(q)=q/2$. Quite remarkably, it is however possible to connect those $\alpha(p)$ to the scaling exponents of the Lagrangian structure function using multi-fractal theory \cite{Biferale}, thereby providing a direct link between scale-symmetry breaking and time-reversal breaking from a Lagrangian point of view.\\
Motivated by such studies, we would like to further explore  Lagrangian intermittency and irreversibility  at different Reynolds number using  recent experimental measurements in a turbulent von K\'{a}rm\'{a}n flow. Eulerian intermittency and irreversibility have already been extensively investigated by our team in such geometry using Particle Image Velocimetry measurements \cite{Saw,Debue1,SawJFM,Dubrulle,Debue21,Faller}.
We have recently performed new campaign of experiments aiming at tracking 40000 particles in a $45 \times 40 \times 6$ mm$^3$ volume with space and time resolution approaching the Kolmogorov length and time. Such measurements are difficult and prone to experimental noise from various sources \cite{Berg}. If successful, they however provide a unique tool to explore turbulence because they also allow to get Eulerian resolved velocity measurements by interpolation onto a fixed grid. Such type of measurements would then prove a unique tool to explore the link between Eulerian and Lagrangian intermittency and irreversibility. \\
In a first step to validate our Lagrangian measurements, we have therefore devised several tools to minimise the noise contribution, both at the level of trajectories \cite{Cheminet21} and structure functions.  We report here the outcome of such effort, and discuss our results by comparing them both to existing literature, and to outcome of the same analysis from trajectories issued from direct numerical simulations (DNS). We also explore how our Lagrangian measurements compare with already existing Eulerian results obtained by our group in the same geometry, and introduce a new tool, the power Lagrangian structure function, that allows to complete the parallel between Lagrangian and Eulerian quantities, from a multi-fractal point of view.

\section{\label{set-up and DNS}Description of the experimental set-up and of the numerical data}
\subsection{Definition of Kolmogorov time and velocity scale}
Kolmogorov quantities (velocity: $u_{K}$ and time scale: $\tau_{K}$) are defined as a function of the viscosity $\nu$ and the energy dissipation rate per unit mass $\epsilon$ as:
\begin{equation}\label{tau_Kolmogorov}
\tau_{K}^G = \sqrt\frac{\nu}{\epsilon},
\end{equation}  
for the time scale, and 
\begin{equation}\label{u_Kolmogorov}
u_{K}^G = (\nu\epsilon)^{1/4}.
\end{equation} 
and for the velocity. 

\subsection{\label{DNS}Description of the Direct Numerical Simulations}
Lagrangian trajectories are computed from a direct numerical simulation of incompressible Navier-Stokes equations. The equations are integrated in a three dimensional box of size $2\Pi$ with periodic boundary conditions using a parallel pseudospectral code with a spatial resolution of $512^3$. Time derivatives are estimated using a second order Runge-Kutta method, and the code uses the 2/3 rule for de-aliasing. As a result, the maximum wave number is $k_{max} = 512/3 = 170$.  A forcing is applied by forcing the wavenumbers ($k_f$=1) associated to a Taylor Green vortex. The Taylor Reynolds number after the transient is $Re_\tau \simeq 150$ which leads to a spatial resolution of  $k_{max} \eta = 1.3$. The simulation is seeded with $128^3$ lagrangian tracers which are integrated in time using a second order Runge-Kutta scheme. The velocity at the tracer's position are obtained at full and half time step using a trilinear interpolation. The position and velocity of the tracers are saved every 5 time steps corresponding to 0.02 Kolmogorov time $\tau_{K}^G$. 

\subsection{\label{set-up}Description of the experimental set-up}
Experiments were performed in a von K\'arm\'an flow generated in a cylinder through counter-rotating impellers fitted with blades. The aspect ratio of the cylinder is 1.8 and its radius ($R$) is 0.1 m. A complete description of the experimental facility can be found in \cite{Saw, Debue1, SawJFM}. \\
The frequency of rotation of the impellers  ($f_{rot}$) is varied to change the global Reynolds number $Re$. It is defined as
\begin{equation}
Re = \frac{2\pi f_{rot}R^{2}}{\nu},
\label{Reynolds}.
\end{equation}
where $\nu$ is the kinematic viscosity of the fluid.
The working fluid used was water. The global energy dissipation in all measurements is constant and equal to $\epsilon=0.045$ when expressed in units of $R$ and $f_{rot}$.\\
The lateral wall of the cylinder was made of plexiglass in order to have optical access. 
The set-up was equipped with four high speed cameras and an high speed laser to perform 4D Particle Tracking Velocimetry (PTV) measurements with the Shake-the-Box algorithm  \cite{Schanz16}. 
Lagrangian tracks were acquired in a volume of about $45 \times 40 \times 6$ mm$^3$ at the center of the cylinder, where turbulence is the most homogeneous and isotropic. Details of the instrumentation and measurement acquisition and processing can be found in \cite{Ostovan}. For each of the cases described in table \ref{exp.program}, we performed 40 independent runs of duration $3226$ $dt$.\\
Experiments were performed at four different Taylor Reynolds numbers, $R_{\lambda}$, by changing the frequency of rotation of the impellers from $f_{rot}=0.1$ Hz to $2.5$ Hz. 
The global $R_{\lambda}^{glob}$ of the experiments was computed from the r.m.s velocities $v_{rms}=\sqrt{{v'}_i^2/3}$ as:
\begin{eqnarray}
R_{\lambda}^{glob} &=&\frac{ v_{rms}\lambda}{\nu},\nonumber\\
\lambda&=&\sqrt{15 \frac{\nu}{\epsilon}}v_{rms},
\label{Reynolds_lambda}
\end{eqnarray}
where $\lambda$ is the Taylor microscale and $v'_i$ is the fluctuation of the ith component of the Lagrangian velocity. Figure \ref{histotraj} shows the computed $R_\lambda$ as a function of $Re$.
We observe a power-law dependence, $R_\lambda\sim Re^{\beta}$, $\beta\approx 0.62$. The exponent $\beta$ is larger than the value $0.5$ expected for homogeneous isotropic turbulence.
We interpret this as a signature of the inhomogeneities and anisotropy of the von K\'{a}rm\'{a}n flow.  In the sequel, we explore the possibility to define a local Taylor Reynolds number, based on local energy dissipation estimates based on trajectories.\\
Table \ref{exp.program} summarises the parameters of the different data sets. 
In the experiments, $dt$ is the inverse of the frequency of acquisition of the high-speed laser.
As one can see from the sixth column of Table \ref{exp.program}, the resolution time of all considered data sets is smaller than $\tau_{K}^{G}$. In particular, in cases A, B and C, the Kolmogorov time is resolved by at least three data points, while in cases D and N by two data points.\\

\begin{table}[h]
\caption{\label{exp.program} 
Summary of the experimental and Direct Numerical Simulations (DNS) parameters.
$Re$ is the global Reynolds number defined in equation (\ref{Reynolds}), which is a function of the characteristic large-scale length and velocity.   
 $(v_{rms}/U)_N$ is the non-dimensional r.m.s. velocity for all trajectories greater than $N$ points. The value for $N=50$ was used to compute the Taylor Reynolds number $R_{\lambda}^{glob}$ using the global measurement of energy dissipation $\epsilon$. 
The term $dt/\tau_{K}^{G}$ represents the ratio between the resolution time, $dt$, and the Kolmogorov time, $\tau_{K}^{G}$, defined in equation (\ref{tau_Kolmogorov}). 
 $N_{10\tau_K}$ is the number of time steps needed to reach the inertial time scale $10$ $\tau_K$. 
 $\epsilon_*$ is the effective local energy dissipation derived from Heisenberg-Yaglom relation for trajectories greater than $N=50$. It is used to compute the local Taylor Reynolds number $R_\lambda$. The last column specifies whether the data were acquired from experiments or DNS.}

\begin{ruledtabular}
\begin{tabular}{cccccccccc}
Case&
$Re$& 
 $(v_{rms}/U)_{5}$ & 
 $(v_{rms}/U)_{50}$ &$R_{\lambda}^{glob}$
&$dt/\tau_{K}^{G}$& $N_{10\tau_K}$ &$\epsilon_*/\epsilon$
&$R_{\lambda}$ &\textrm{Source}\\
\colrule
A&$6.3 \times 10^3$ &0.41 &0.37 &64 & 0.06 &166 &0.66 &81 &Experiment\\
B&$3.1\times 10^4$ &0.51 &0.41  &184 & 0.15 &67 &0.58 &248 &Experiment\\
C&$6.3 \times 10^4$ &0.53 &0.48  &327 & 0.22 &45 &0.49 &507 &Experiment\\
D&$1.6 \times 10^5$ &0.54 &0.43 &444 & 0.34 &29 &0.37 &762 &Experiment\\
N&652 & & &152 & 0.38  &26 & & &DNS\\
\end{tabular}
\end{ruledtabular}
\end{table}

\subsection{Description of the data sets}
\begin{figure}
(a)\includegraphics[width=0.47\textwidth]{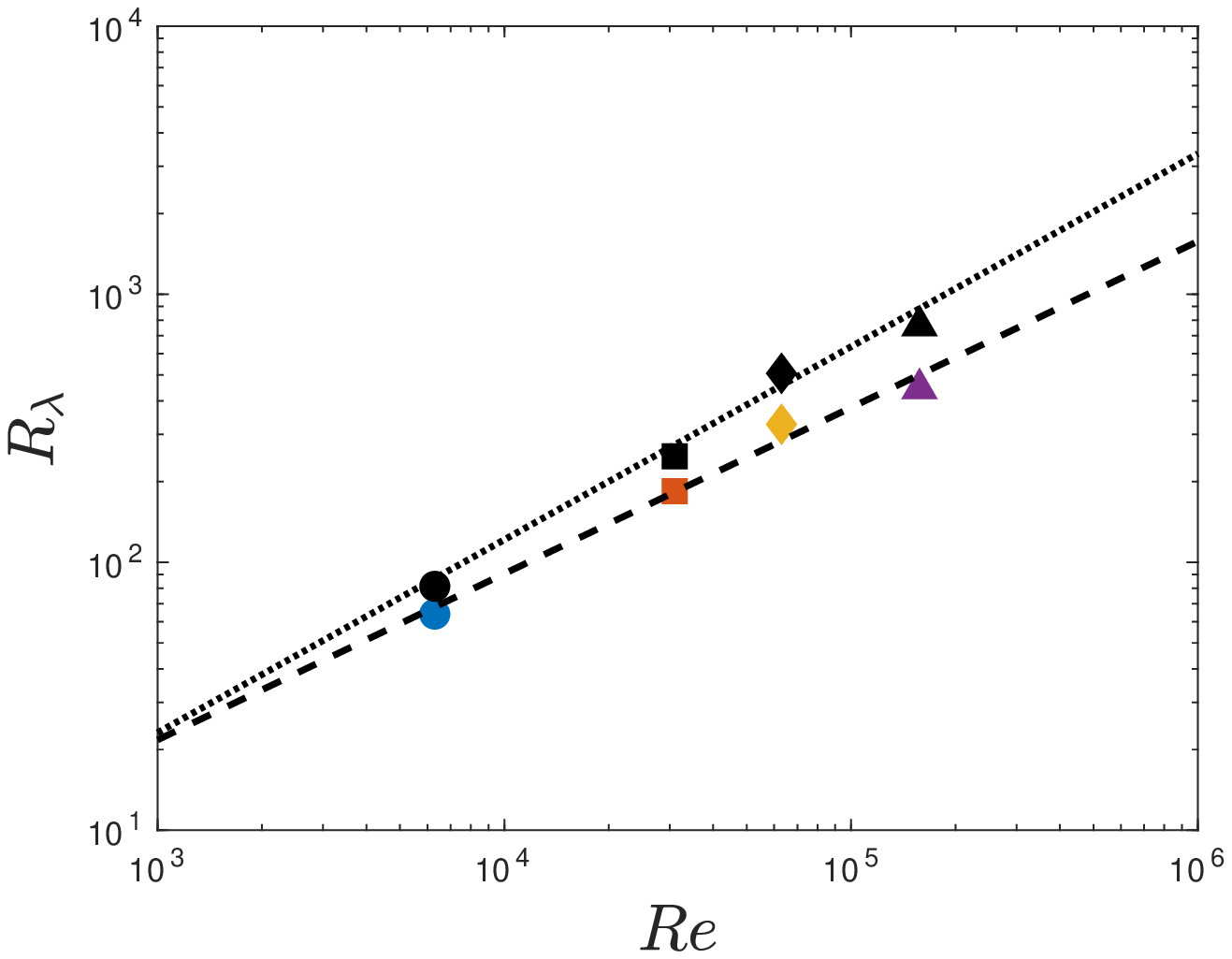}
(b)\includegraphics[width=0.47\textwidth]{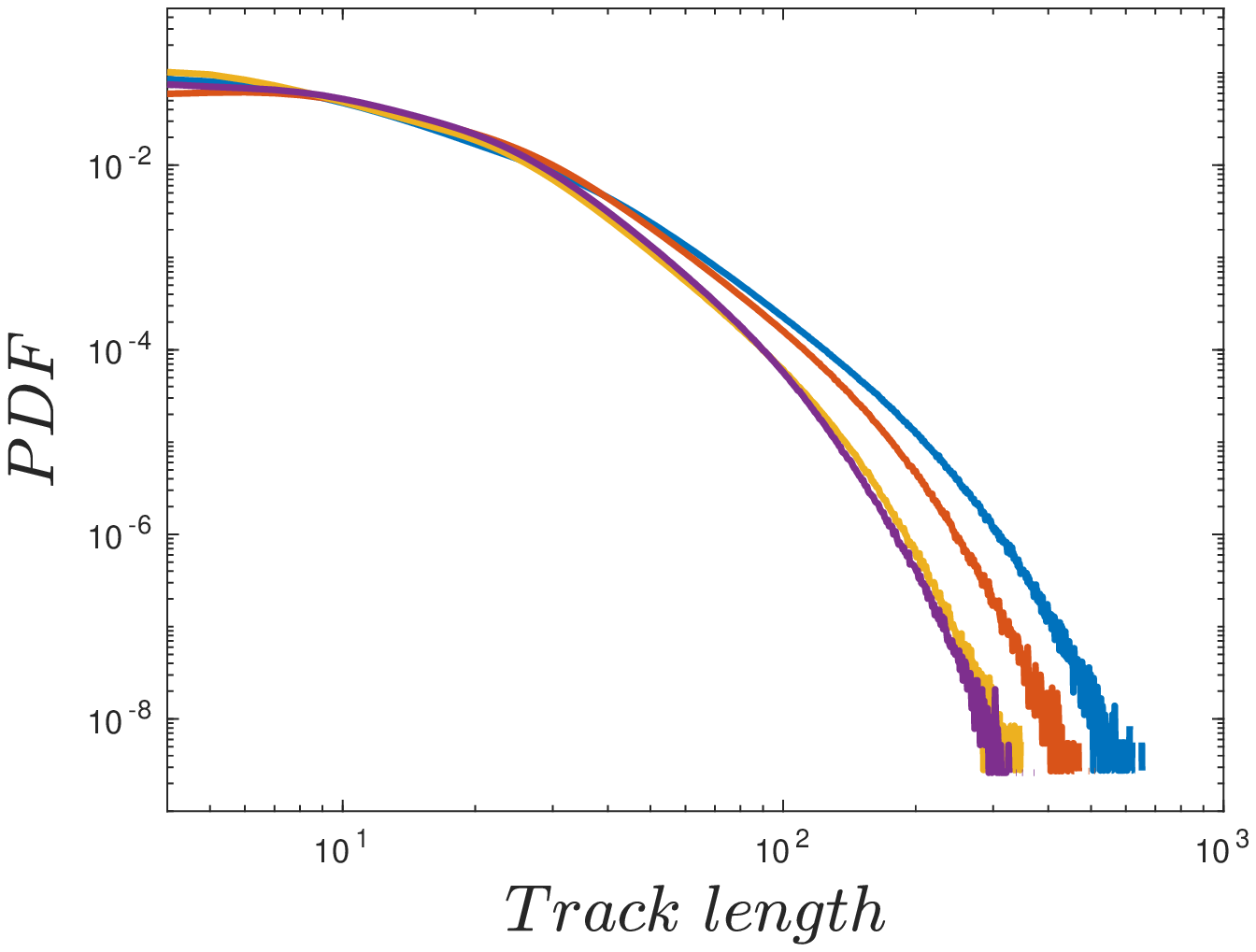}
\caption{(a) Experimental Taylor Reynolds number ($R_\lambda$) vs global Reynolds number ($Re$) for experiments A (blue circle), B (red square), C (yellow diamond) and D (magenta triangle). The black symbols are the Reynolds numbers corrected by the local dissipation $\epsilon_*$. The dashed line is a power law $R_\lambda=0.3 Re^{0.62}$ and the dotted line $R_\lambda=0.16 Re^{0.72}$. (b) PDFs of trajectory lengths obtained for case A, B, C, and D. Same colours as in panel (a).}
\label{histotraj}
\end{figure}

The tracking algorithm provides us with a set of $N=O(10^7)$ experimental trajectories $\vec{r}(t)$ per run, for each of the cases of table \ref{exp.program}. A typical histogram of the length of the trajectories is shown in figure \ref{histotraj} for all experimental cases. We show that a vast majority of the trajectories have a length less than 100 $dt$, which means that our measurements are not well fitted to describe the inertial range $\tau>10$ $\tau_K$, except for case D. This highlights a first difficulty of this type of measurements.

In addition, we have many very small trajectories of length less than 50.  These trajectories contribute significantly to the value of  the r.m.s. of velocity, as can be seen in table 1. We  found that such trajectories are the noisiest ones (see below for a check of their influence), and we discarded them from all our analysis, leaving 
$O(10^8)$ trajectories over 40 runs per case for our analysis. This is sufficient to converge moments of velocity increments up to $6$ for small increments, but for large increments, which require 
longer and less numerous trajectories, our statistics starts to deteriorate from order 5.\\
In the numerical case, we have $N=10^5$ trajectories of length $800$ $dt$, with no noise. We therefore kept all trajectories.


\section{\label{analysis}Data analysis methods}
The same data analysis method described in paragraph \ref{filter} is applied to treat both the experimental and the DNS data. This method is then used to extract the Lagrangian physical quantities described in paragraph \ref{LagrangianPower} and \ref{definitions}. The comparison between the results from the experiments and from the DNS data at similar $Re_{\lambda}$ allowed to check the influence of the noise on the results, and the ability of our treatments to deal with it. 

\subsection{\label{filter}TrackFit filtering and coarse-graining through scales}
Experimental measurements issued from the tracking algorithm  are usually polluted by noise, which makes them difficult to differentiate with respect to time to get velocity and acceleration. 
For this reason, these quantities are usually only computed after some noise filtering has been executed. The easiest procedure consists in applying a low-pass filter on the trajectories. For instance, this can be done by convoluting the trajectories with the first or second derivative of the filter's kernel to get the velocity or acceleration along the trajectories. The practical issue is then to choose the width of the filter, to remove as much noise as possible without getting rid of physical signal.\\ 
In the present paper, we tested two methods to compute the velocity structure functions.
 \subsubsection{Direct method}
The first one, is to use our in-house trajectory smoothing method, named TrackFit \cite{Gesemann16} based on regularised B-spline. In \cite{Cheminet21}, we improved on Gesemann's original work to select the best filter lengthscale. The idea is to set the TrackFit's cutoff frequency $f_c$ at the frequency at which the Signal-to-Noise-Ratio (SNR) is equal to one. This can be an arduous task when confronted to real non-white experimental noise. A criterion based on the spectral behaviour of the  residual vector was found to give minimal error on a large range of scales when using this algorithm. The filtering operation was done on the three components of the trajectories.\\
The final Lagrangian velocity resolution is given by the chosen filter cut-off frequency. Below this frequency, measured waves are considered as real physical signal. Above this frequency, measured waves are considered to be noise and thus are filtered out and their amplitudes are damped following a $f^{-6}$ law. The velocity is thus regularised for timescales below the filter's timescale $\Delta_c = 1/(2f_c)$.  $\Delta_c$ is estimated at $0.7$ $\tau_{K}$, $1$ $\tau_{K}$, $1.2$ $\tau_{K}$, and $1.8$ $\tau_{K}$ for respectively case A, B, C, and D.\\ 
The algorithm then provides us with the values of velocity $v=dr/dt$ and acceleration $a=d^2r/dt^2$ along the trajectories, from which we can compute the velocity increments $\delta_\tau v= v(t+\tau)-v(t)$ and the instantaneous power along the trajectory :
\begin{equation}\label{P}
P=\vec{u}\cdot D_{t}\vec{u}=\vec{u}\cdot \vec{a}.
\end{equation}

From these quantities, we define the Lagrangian velocity structure functions of order $p$  as:
\begin{equation}\label{Sp}
S_{p}(\tau)=\langle((\delta_\tau v)^{p}\rangle,
\end{equation}  
where the operator $\langle \cdot \rangle$ represent the ensemble average over all the trajectories. In the K41 framework, we expect $S_{p}(\tau)\sim (\epsilon \tau)^{p/2}$. On the other hand, 
if the velocity field is regular, then $S_{p}(\tau)\sim \tau^{p}$. This scaling is expected for very small $\tau$, in the absence of any singularity.\\

Using the Lagrangian structure function, we may also compute the flatness of the Lagrangian velocity, defined as:
\begin{equation}\label{eq_flatness}
F = \frac{S_{4}(\tau)}{[S_{2}(\tau)]^{2}}
\end{equation}

We also define Lagrangian power moments as:
\begin{equation}\label{Sp_P}
R_{p}(R_\lambda)=\langle\vert P\vert ^p\rangle.
\end{equation}  
Inspired by  \cite{Jucha} and \cite{Biferale}, we also define an irreversibility indicator by taking the signed moments of the Lagrangian power:
\begin{equation}\label{Ap_P}
A_{2p+1}(R_\lambda)=\langle\left(P\right)^{2p+1}\rangle.
\end{equation} 
If turbulence was invariant under time reversal, any odd moment should be zero. Therefore they are the hallmark of irreversibility. 

 \subsubsection{Wavelet increments}
 A second method uses an adaptation of the Eulerian wavelet velocity increments (WVI) to Lagrangian measurements. In this procedure, we start from trajectories that have been 
 subject to only a preliminary filtering and use  a coarse-graining procedure at a characteristic scale $\tau$, that only applies onto the trajectories $\vec{r}$. Specifically, we  convolute them with derivatives of a filtering function , $\phi$, which is smooth, non negative and with unit integral, $\int\phi^\tau (\xi)d\xi =1$. This has a Gaussian shape: $\phi^\tau(\xi)\propto e^{-\xi^2/2\tau}$. We call such procedure "weak derivative"  \cite{Leray}. The advantage of such procedure is that it is faster than the direct method, because it spares the full TrackFit optimisation that is rather costly, and replace it with wavelet transform, that can be made very fast using Fast Fourier Transform \cite{Dubrulle}.\\

\par {\bf The time-smoothing of a trajectory at scale} $\tau$ is expressed by:
\begin{equation}\label{rtau}
r_{i}^{\tau}(t)=\int_{-\infty}^{\infty}\phi^{\tau}(\xi)\vec{r}(t+\xi)d\xi.
\end{equation}\\
{\bf Velocities ($\vec{v}$) and accelerations ($\vec{a}$) at scale $\tau$} were both computed directly from the raw trajectories using ''weak derivatives'' of first and second order:
\begin{equation}\label{vtau}
v_i{}^{\tau}(t)=\int_{-\infty}^{\infty}\nabla\phi^{\tau}(\xi)\vec{r}(t+\xi)d\xi,
\end{equation} and
\begin{equation}\label{atau}
a_{i}^{\tau}(t)=\int_{-\infty}^{\infty}\Delta\phi^{\tau}(\xi)\vec{r}(t+\xi)d\xi, 
\end{equation}
with $i=X,Y,Z$, and where the gradient ($\nabla$) and the Laplacian ($\Delta$) operators are applied to the filtering function and not to the trajectory. 
When the scale tends to zero, the filtered quantities in equations (\ref{rtau}), (\ref{vtau}), and (\ref{atau}) tend to their correspondent unfiltered quantity, i.e.:
\begin{equation}\label{lim_rtau}
\lim_{\tau \to 0} r_{i}^{\tau}(t)=r_{i}(t), \lim_{\tau \to 0} v_{i}^{\tau}(t)=v_{i}(t), \lim_{\tau \to 0} a_{i}^{\tau}(t)=a_{i}(t).
\end{equation}\\
The convolution integrals were computed with continuous wavelets transforms ($WT$s), based on fast Fourier transforms. Before computing the weak derivatives, the trajectories were extended by mirroring half of them on both sides. In this way, the extended trajectories are periodic, and no noise is added by the fast Fourier transform operation. After computing the derivatives, only the points corresponding to the original trajectories were taken.
At the extremes of the trajectories the weak derivatives are not accurate for a number of points that is proportional to the support of the filtering function used. 
In this case the length of the inaccurate regions is proportional to the filter scale, and a length of three times the width of the support of the filtering function was cut from the filtered data. This provides an additional limitation of 
the largest $\tau$ we can reach with our data, in the case of wavelet velocity increments. For example, for a trajectory of length $100$ $dt$, subject to a filter with width $10$ $dt$, we have to remove $30$ points on each side, leaving only $40$ points to provide statistics for the wavelet increments, and we cannot go beyond $16$  $dt$ for such trajectory. This means that we can sample smaller time increments than the direct method. In particular, this method is not well fitted to access inertial range properties in our data, given the small number of very long trajectories.
\subsection{\label{LagrangianPower}The Lagrangian power at scale $\tau$}
We would like to mirror what we did for the velocity and acceleration, and define a "Lagrangian power at time $\tau$", obtained by smooth derivative of a quantity based only on particle positions over the time increment $\tau$ ($D_\tau$). 
For this, we notice that 
\begin{eqnarray}
D^3_\tau (r(t+\tau)-r(t))^2&=&6 D_\tau r(t+\tau) D_\tau^2 r(t+\tau),\nonumber\\
&+&2  D_\tau^3 r(t+\tau)\left(r(t+\tau)-r(t)\right).\nonumber
\end{eqnarray}
In the limit $\tau\to 0$, $D_\tau r(t+\tau) \to v(t)$, $D_\tau^2 r(t+\tau)\to a(t)$, and $r(t+\tau)-r(t))\to 0$.
Therefore, the quantity:
\begin{equation}
\Pi^{\tau}(t)=\sum_{i=X,Y,Z}\frac{1}{6}\int_{0}^{1}{D_{t}^{3}\phi(\xi)\left[r_{i}(t+\xi)-r_{i}(t)\right]^2 d\xi}
\end{equation}
is such that for scale $\tau$ that tends to zero:
\begin{equation}
\lim_{\tau \to 0} \Pi^{\tau}(t)=P(t).
\end{equation}

The quantity $\Pi^{\tau}$ can therefore be seen as a coarse-grained version of $P$ at scale $\tau$. The Kolmogorov theory K41 combined with a Taylor hypothesis predicts
that the velocity increments $v(t+\tau)-v(t)$ scale like $(\epsilon \tau)^{1/2}$ in some inertial range. Integrating with respect to time, we thus expect that $r(t+\tau)-r(t)\sim\epsilon^{1/2} \tau^{3/2}$. So, if both K41 and local Taylor hypothesis are right, we expect that 
\begin{equation}
\langle \Pi^{\tau}\rangle \sim \epsilon,
\label{pseudoRich}
\end{equation}
 Intermittency will then be detected via deviations from this law, in a way analogous to the K62 phenomenology. Note that equation (\ref{pseudoRich}) can also be viewed as a kind of Richardson law 
 along the trajectories, as it implies that for large values of $\tau$, $\langle (r(t+\tau)-r(t))^2\rangle \sim \epsilon \tau^3$ along trajectories.\\

\subsection{Wavelet structure functions}
\label{definitions}
Lagrangian wavelet  structure functions of order $p$ are defined as:
\begin{equation}\label{Wp}
W_{p}(\tau)=\langle(\mid{\vec {a}^{\tau}}\mid\cdot\tau)^{p}\rangle,
\end{equation}  
where the operator $\langle \cdot \rangle$ represent the ensemble average over all the trajectories. They should have the same scaling property than the Lagrangian velocity structure functions, namely scale like  $W_{p}(\tau)\sim \tau^{p}$ at small $\tau$ and scale like  $W_{p}(\tau)\sim (\epsilon \tau)^{\zeta(p)}$ in the inertial range, where $\zeta(p)=p/2$ if K41 holds.\\
Using $\Pi^{\tau}$, we may define Lagrangian power structure functions as:
\begin{equation}\label{Sp_P}
T_{p}(\tau)=\langle\mid{\Pi^{\tau}}\mid^{p}\rangle.
\end{equation}  
In the K41 framework, we expect $T_{p}(\tau)\sim (\epsilon)^{p}$.

\section{\label{results}Results}
\subsection{Local energy dissipation}
\label{locediss}
The von K\'{a}rm\'{a}n flow being non-homogeneous, the local dissipation $\epsilon$ at the location of measurements may differ from its global value, measured in the whole tank  in a manner that depends on the Reynolds number (see e.g. discussion in \cite{Saw}).  To try to estimate such local dissipation, we may use Heisenberg-Yaglom relation, that connects the acceleration variance $<a^2>$, $\epsilon$ and the Taylor Reynolds number $R_\lambda$ through 
\begin{equation}
\frac{<a^2>}{(\epsilon^3/\nu)^{1/2}}=1.9\frac{R_\lambda^{0.135}}{1+(85/R_\lambda^{1.135})},
\label{YaglomHeisenberg}
\end{equation}

\cite{Heisenberg1948, Yaglom1949, Lawson18}. To illustrate this, we plot in figure \ref{fig_Yag}(a)  the non-dimensional acceleration $<a^2>/(\epsilon^3/\nu)^{(1/2)}$ as a function of $R_\lambda$, both computed using
$\epsilon$ measured globally through torques \cite{Brice}. We further separate the trajectories in two categories: those with length $N$  greater than $N=5$ and those with length greater than $N=50$, which are less prone to experimental noise. We see that in the latter case, all points are significantly smaller than the line representing the Heisenberg-Yaglom relation, while in the former case, only the points corresponding to the two largest Reynolds numbers are below the line. This could be corrected by using a smaller value of $\epsilon$, meaning that the local dissipation in our measurements is smaller than the global dissipation. This is due to two causes: first, non-homogeneity. If this was the only cause, we should however not expect a difference between the two selections of trajectories. In fact, we are likely to observe the influence of two additional biases: one due to the fact that the experimental noise increases the variances of both velocity and acceleration. This tends to shift the points to the upper left part of the diagram, as observed. There is however an additional physical bias which is due to the fact that less noisy trajectories are expected in places where the flow has a milder behaviour (lower velocities, less irregularities), implying a lower dissipation. By selecting trajectories with $N \geq 50$, we therefore bias our measurements towards regions of lower dissipation, and , presumably, lower intermittency. This possibility will be explored in the sequel.\\
We have estimated what is the value of $\epsilon$ that should be used so that our measurements for $N \geq 5$ collapse on the Heisenberg-Yaglom line. The corresponding $\epsilon_*$ is provided in table \ref{exp.program}. The corresponding estimate of $\epsilon_*$ can then be used to redefine a Kolmogorov velocity and time that is hopefully cleaned with respect to inhomogeneity and anisotropy and trajectory length selection bias. We did not do the same thing for the case $N \geq 5$ because of noises issues that are explored in section \ref{noiseissue}. We also recomputed the local Taylor Reynolds number
using the local dissipation and plot it on figure \ref{histotraj}. It follows a steeper power law a function of the Reynolds number $Re$, with exponent close to $0.72$.

\subsection{Influence of trajectory length on the structure functions}
\label{noiseissue}
\begin{figure}
(a)\includegraphics[width=0.47\textwidth]{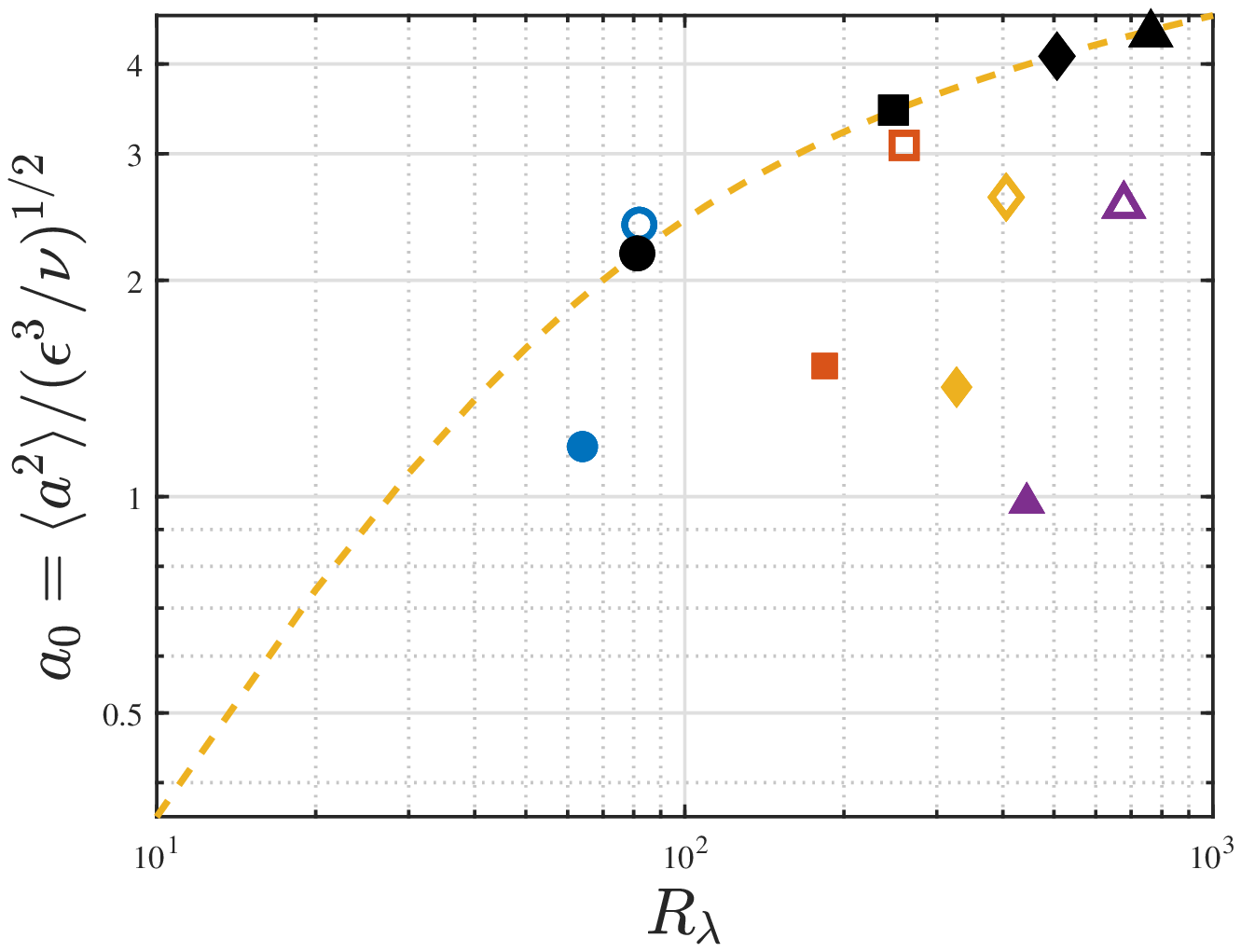}
(b)\includegraphics[width=0.47\textwidth]{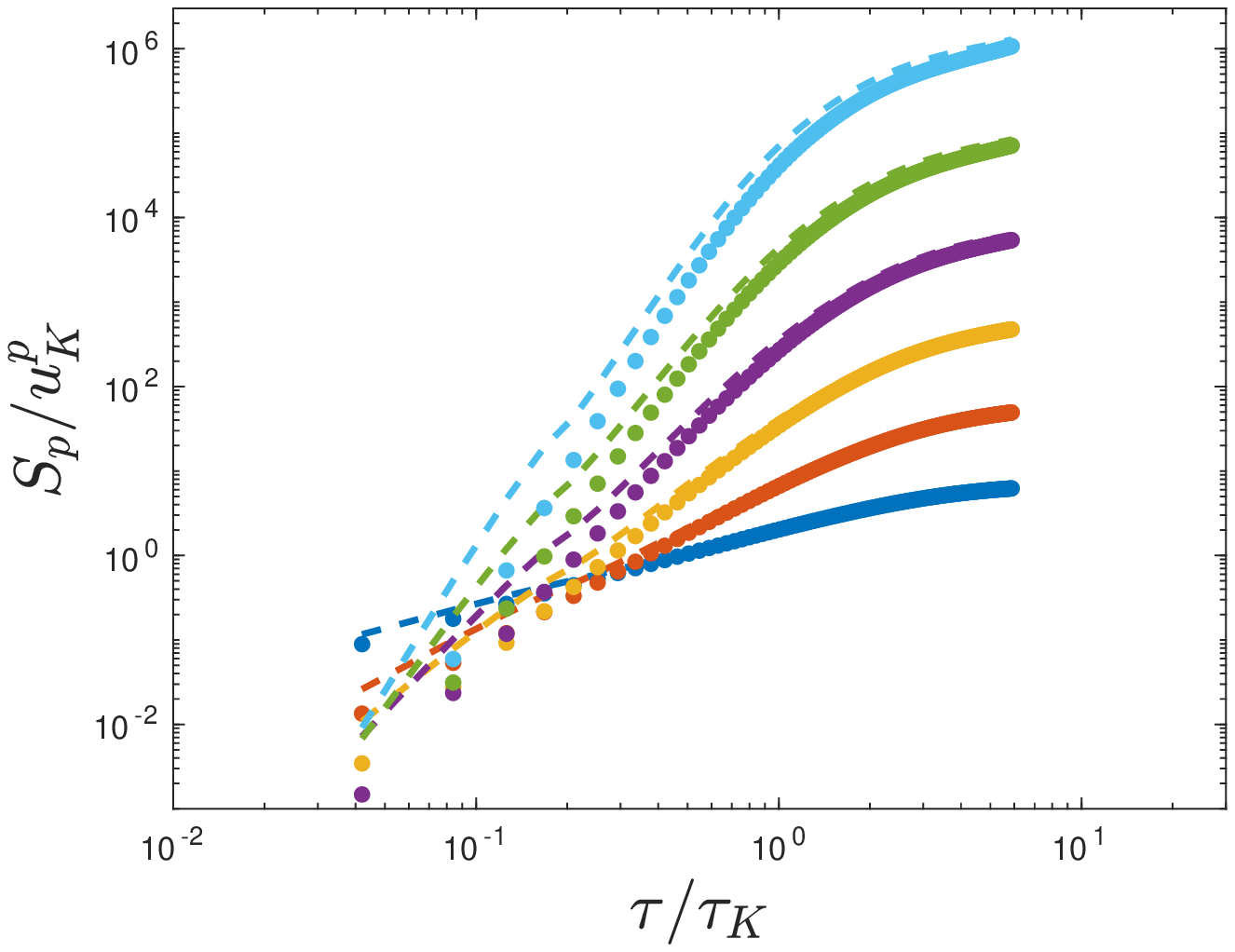}
\caption{(a) Comparison with the Heisenberg-Yaglom relation $\frac{<a^2>}{(\epsilon^3/\nu)^{1/2}}=1.9\frac{R_\lambda^{0.135}}{1+(85/R_\lambda^{1.135})}$ (yellow dotted line) for case A (circle), B (square), C (diamond), and D (triangle).
Open symbols represent trajectories with length greater than $5$ $dt$, while filled symbols trajectories with length greater or equal than $50$ $dt$. The coloured symbols are the raw data, computed using the global dissipation $\epsilon$. The black symbols are the corrected data, computed using the local dissipation $\epsilon_*$ . (b) Influence on trajectory length on Lagrangian velocity structure functions of order $p = [1:6]$ (blue, red, yellow, magenta, green, cyan, respectively), scaled by local Kolmogorov quantities, for case A. Trajectories with length larger than $50$ $dt$ are shown by circles and trajectories larger than $5$ $dt$ by dashed lines.}
\label{fig_Yag}
\end{figure}

As previously said, we have observed that trajectories with length smaller than $N=50$ $dt$ are more noisy. The influence of such noise can be seen by computing the structure functions for 
all trajectories with length larger than $5$ $dt$ and larger than $50$ $dt$. The comparison is done in figure \ref{fig_Yag}(b) for case A (the other cases show similar effects). We see that the structure functions computed using smaller trajectories display a bump at small $\tau$, characteristic of the noise influence. At larger $\tau$ the difference between the two selections decreases, since the short trajectories are less and less taken into account in the computation. In the sequel, we shall then only focus on the structure functions computed using trajectories larger than $50$ $dt$.

\subsection{Velocity structure functions scaling}
\begin{figure}
(a)\includegraphics[width=0.47\textwidth]{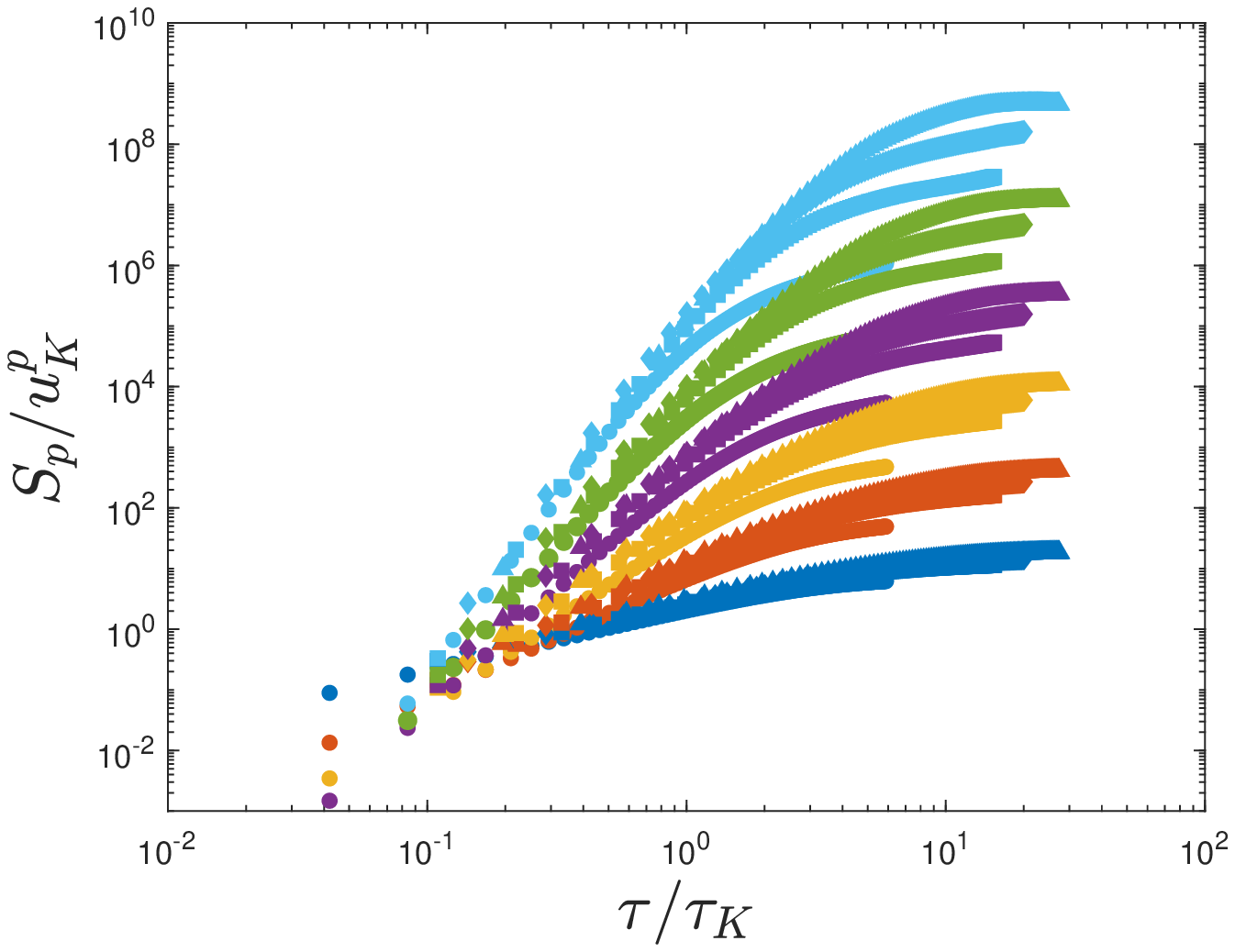}
(b)\includegraphics[width=0.47\textwidth]{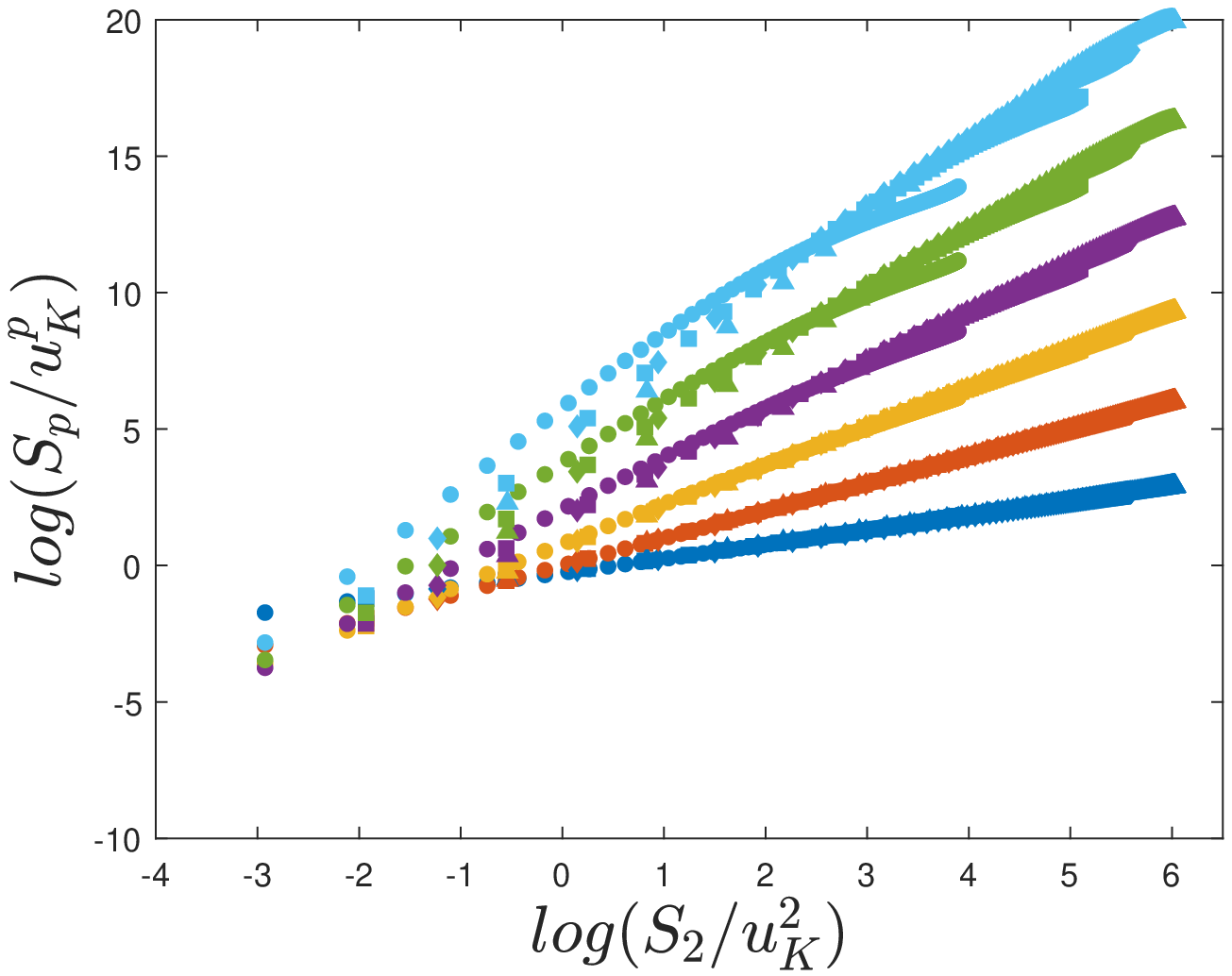}
\caption{Lagrangian velocity structure functions of order $p = [1 : 6]$ (blue, red, yellow, magenta, green, and cyan, respectively) for case A (circle), B(square), C(diamond), and D(triangle), as a function of:  $\tau/\tau_K$, where $\tau_{K}$ is the local Kolmogorov time, in panel (a) and of $S_2$ (ESS property) in panel (b) $u_{K}$ is the local Kolmogorov velocity.\label{fig_velSp}}
\end{figure}

Figure \ref{fig_velSp} shows the Lagrangian velocity structure functions defined in equation (\ref{Sp}) 
as a function of time, scaled by the local Kolmogorov quantities, computed using $\epsilon_*$ rather than $\epsilon$.
A rather good collapse onto a line parallel to the laminar prediction $\tau^p$ can be observed for experimental data at small scale $\tau$  all Reynolds numbers, reminiscent of the universality observed for Eulerian velocity structure functions in that geometry \cite{Saw,Dubrulle}. 

For larger scale $\tau$, the structure functions gradually bend towards another scaling regime.
As already observed in the Eulerian context, such scaling regime is not well defined when looking at the scaling as a function of $\tau$, but its is better defined when defined as a function of, say, the second order structure function $S_2$ \cite{Mordant} (ESS property \cite{Benzi}). This is illustrated in figure \ref{fig_velSp}(b), where clear scaling laws are observed, that can be fitted by a relative exponent $\zeta(p)/\zeta(2)$ that is reported in table  \ref{exp.program}. Such values are higher than the values reported in a similar von K\'{a}rm\'{a}n geometry by Mordant et al. 2004 at $Re_{\lambda} = 510$ \cite{Mordant}, meaning that there is less intermittency.

\begin{figure}
(a)\includegraphics[width=0.47\textwidth]{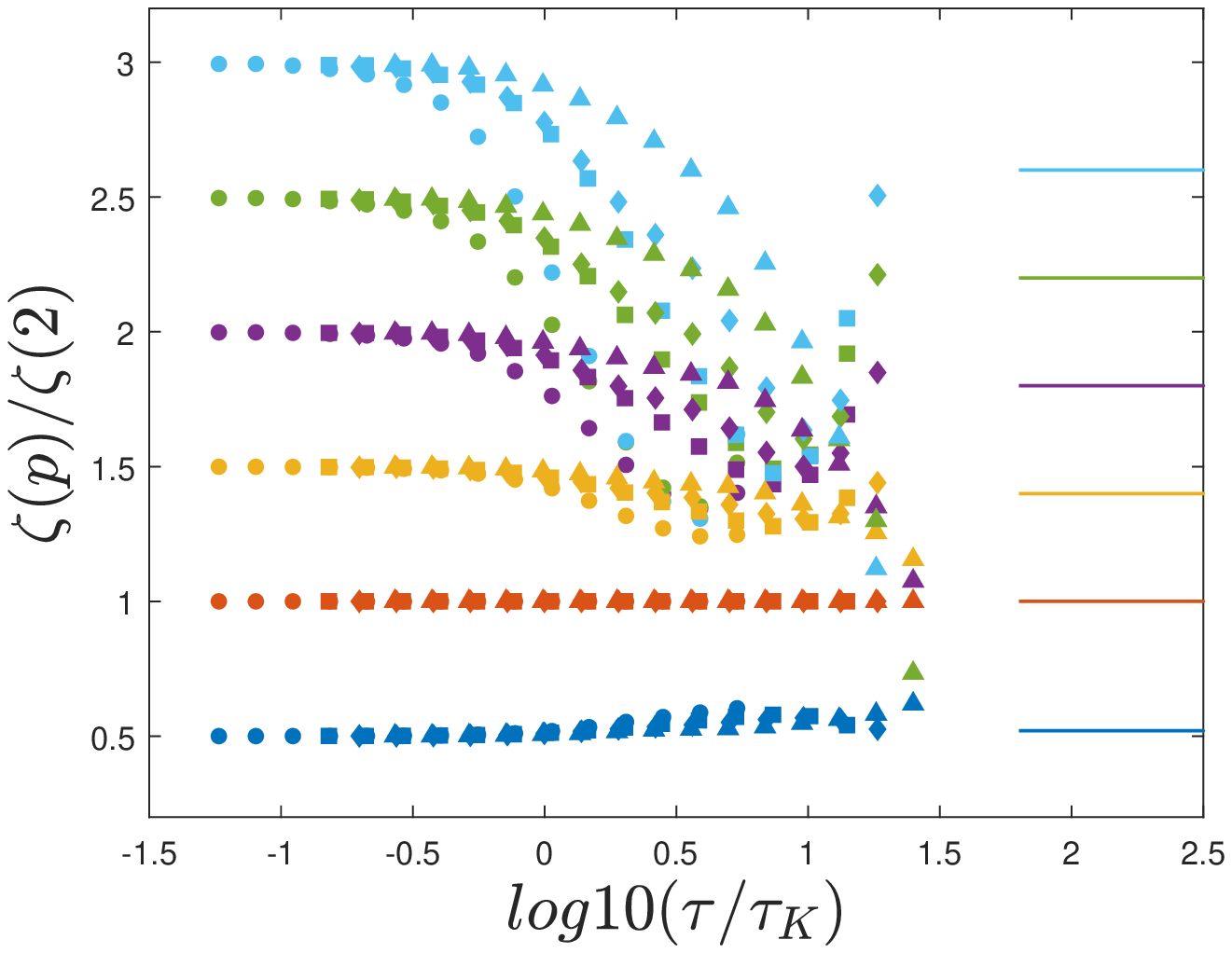}
(b)\includegraphics[width=0.47\textwidth]{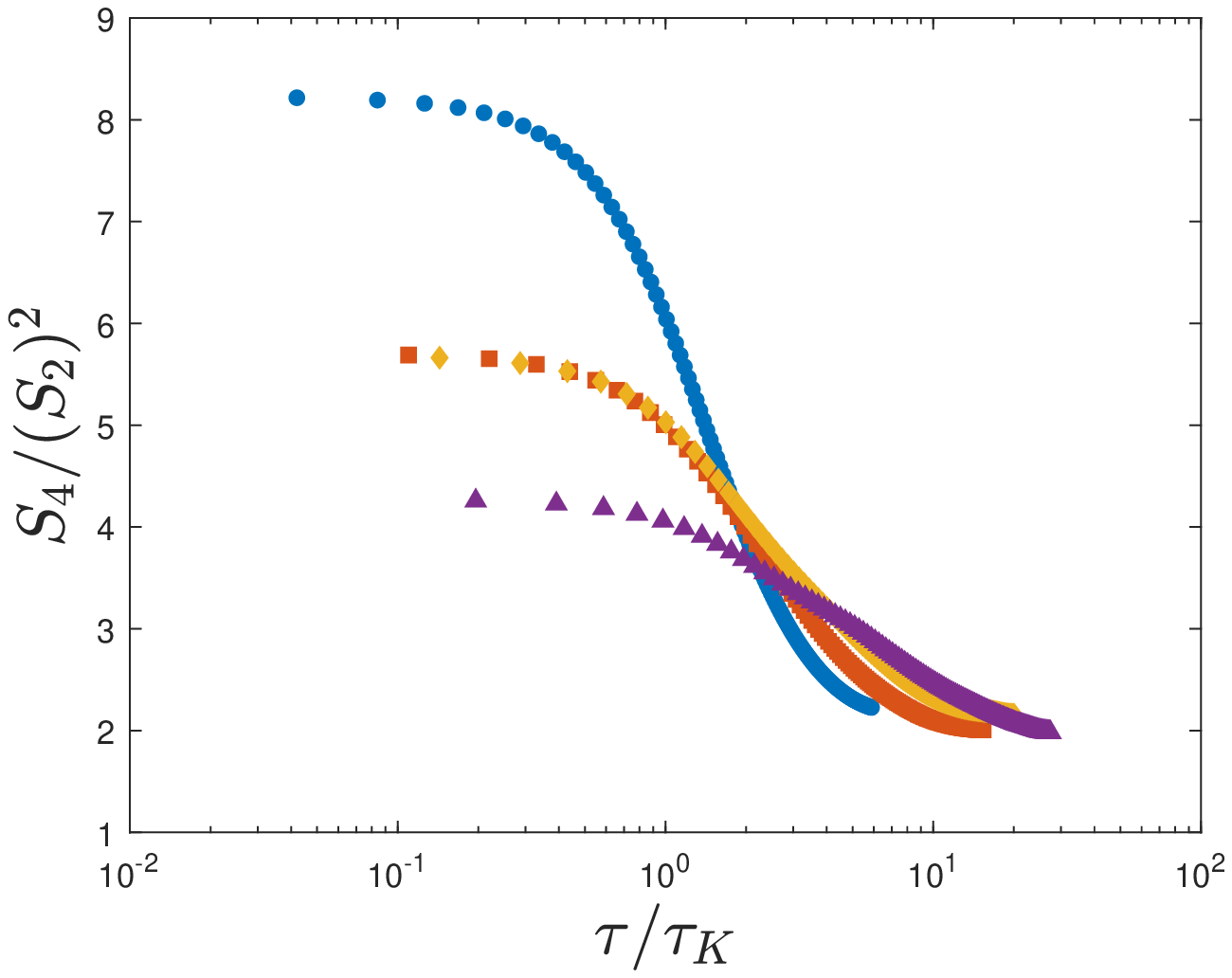}
\caption{(a) Local scaling exponents of orders $p = [1 : 6]$ (blue, red, yellow, magenta, green, cyan, respectively) for cases A (circle), B (square), C (diamond), and D (triangle) as a function of $\tau/\tau_{K}$, where $\tau_{K}$ is the local Kolmogorov time. The five horizontal lines on the right indicate the values obtained through fits using the ESS property. (b) Flatness for cases A (blue circles), B (red squares), C (yellow diamonds), and D (magenta triangles) as a function of $\tau/\tau_{K}$.\label{fig_expoSp}}
\end{figure}

We note however that the improvement of the scaling is in part illusionary, as can be checked by computing the local derivative $\zeta(p,\tau)d\ln S_p/d\ln\tau$ and use it to define a local relative scaling exponent $\zeta(p,\tau)/\zeta(2,\tau)$ shown in figure \ref{fig_expoSp}. On such plot, we can observe clearly a transition towards regularity for small values of $\tau$, where all local exponent converges towards the value $p/2$. This feature is observed both in experiments and in the DNS data.  However, at larger values of $\tau$, there is a transition towards a regime where the local exponent is smaller, but  scatter/oscillate around the values computed using ESS, that are reported by horizontal lines. The scatter is small for orders less than 3, but becomes larger for higher orders, due probably in part to deteriorating statistics. 


\subsubsection{Comparison with multi-fractal theory}

Under the K41 scaling, $\zeta(p)=p/2$ so that relative exponents 
should also scale like $p/2$. To check this, we plot in figure \ref{fig:compaMFR} the relative value obtained through ESS as a function of order. As can be seen, there is a discernible deviation from the linear behaviour (blue continuous line), characteristic of a multifractal (MFR)  behaviour. As discussed in (Chevillard \cite{Chevillard,  Schmitt, Kamps}), the prediction from the MFR theory for homogeneous isotropic turbulence is that:
\begin{equation}
\zeta(p)=\min_{h}\left(\frac{ph+C(h)}{1-h}\right),
\label{predictionMFR}
\end{equation}
where $C(h)$ is the same function than what is measured with the Eulerian velocity structure functions \cite{Chevillard, Schmitt, Kamps}. In the homogeneous isotropic case, a good model of $C(h)$ is a log-normal model $C(h)=(h-a)^2/2b$, with $a=1/3+3b/2$ and $b=0.025$ \cite{arneodo}.   

In the von K\'{a}rm\'{a}n case, the log-normal model also provides a fair fit of the Eulerian structure function, provided  $a=0.35$ and $b=0.045$ \cite{Dubrulle}. Plugging this function into eq. (\ref{predictionMFR}), we obtained multifractal prediction for $\zeta(p)/\zeta(2)$ that is reported on Figure \ref{fig:compaMFR}. We see that it predicts values of exponents that are close to the observations by \cite{Mordant} that were performed in a von K\'{a}rm\'{a}n set up with different propellers. However, it does underestimate strongly the value that we observe, that are better fitted by a log-normal model with $a=0.4$ and $b=0.015$. This is characteristic of a much lower intermittency, and could be due to the fact that we bias our measurements through areas of higher regularity by selecting long trajectories (see section \ref{locediss});

\subsubsection{Comparison with Lagrangian wavelet structure function}
\begin{figure}
(a)\includegraphics[width=0.47\textwidth]{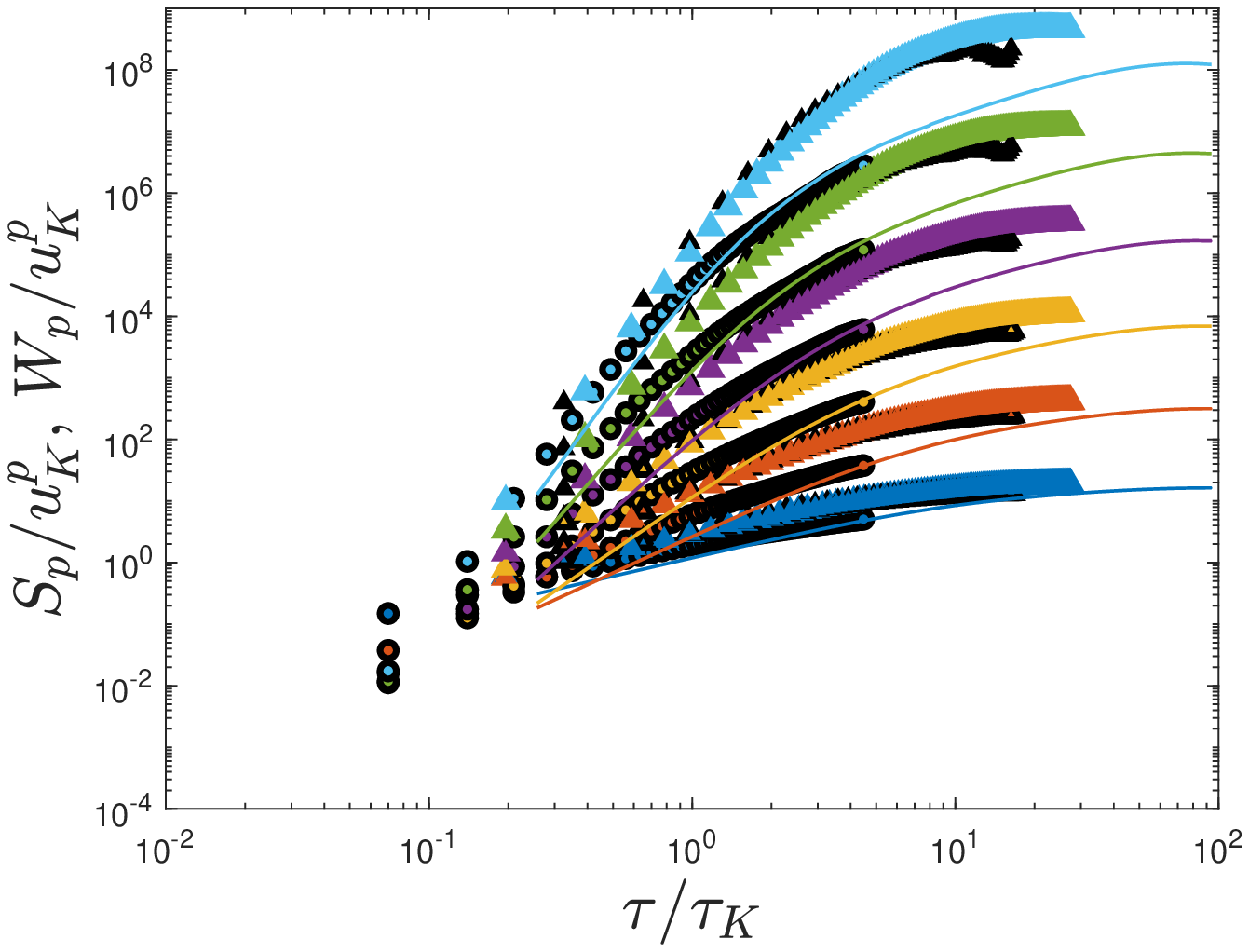}
(b)\includegraphics[width=0.47\textwidth]{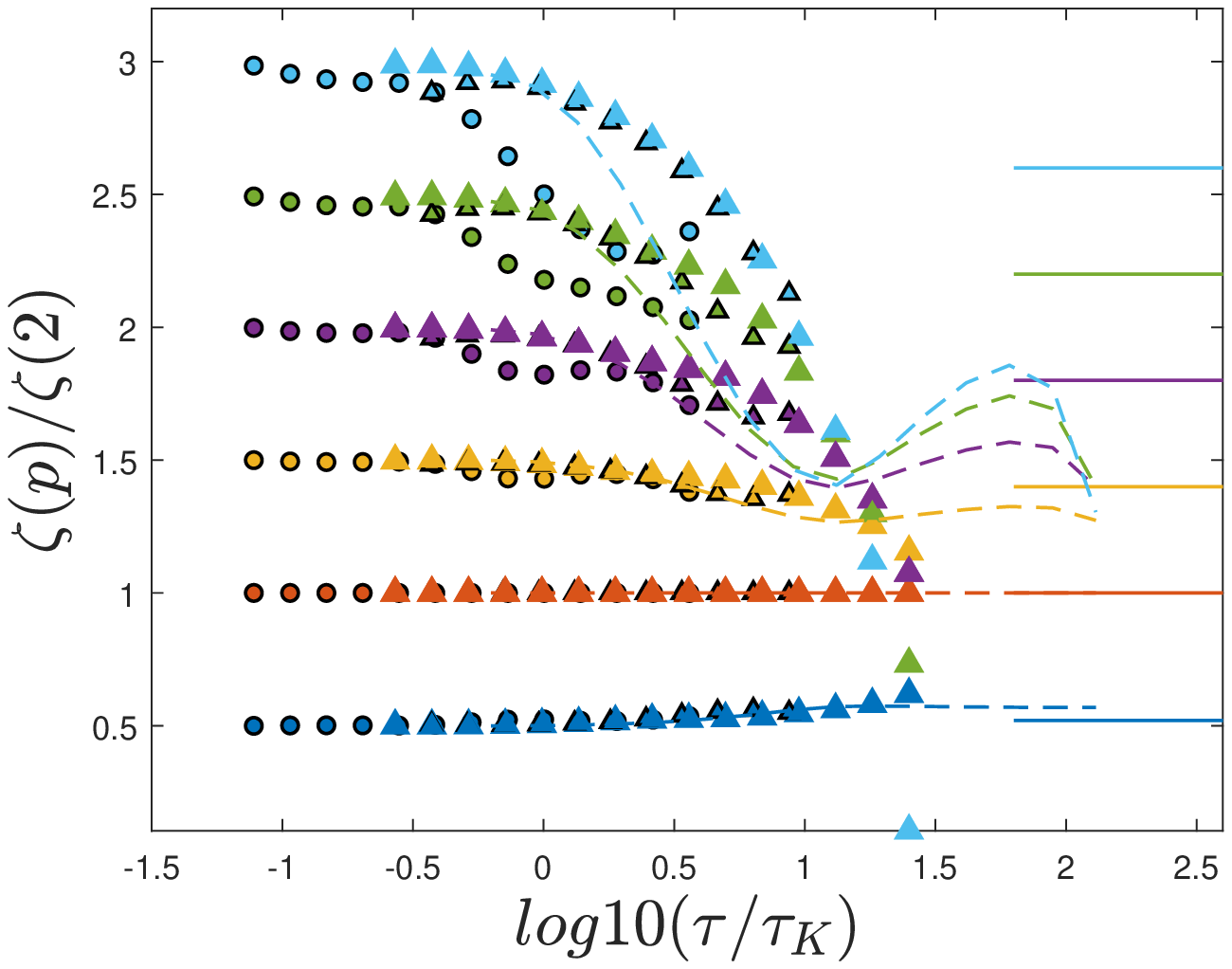}
\caption{(a) Comparison between Lagrangian velocity structure functions and wavelet structure functions (filled symbols without and with black perimeter, reciprocally) of order $p = [1 : 6]$ (blue, red, yellow, magenta, green, and cyan, respectivelly), for cases A (circle), D (triangle), and N (lines), as a function of $\tau/\tau_{K}$, where $\tau_{K}$ and $u_{K}$ are the local Kolmogorov time and velocity, respectively. (b) Same comparison as in panel (a) for local relative scaling exponents. In this plot case N is shown by dashed lines. The five horizontal lines on the right indicate the values obtained through fits using the ESS property. }\label{fig_compavelSp}
\end{figure}

We further computed the Lagrangian wavelet structure functions, starting from a set of data with no prior optimised noise removal. The result is shown on figure \ref{fig_compavelSp}(a) for the cases A, D, and N. We see that globally, the wavelet structure functions agree with the velocity structure functions for all orders. However, differences appear when one looks at the local exponents, shown in figure \ref{fig_compavelSp}(b): it appears more noisy in the wavelet case, and allows to reach a smaller range of scales. This is an indication that wavelet filtering is not sufficient to remove the small scale temporal noise present in the data, and highlights the interest of our TrackFit optimisation, for exploring the small frequencies. At inertial time scale, the wavelet method provides the same estimate of the local scaling exponent than the direct method, a feature that was also observed for Eulerian structure functions \cite{arneodo}. To check our algorithm, we also apply it to the trajectories issues from the DNS. We observe that the local exponent seems to saturate in the inertial range towards a smaller exponent than our experimental measurements meaning more intermittency. Indeed, we did not introduce any bias in the DNS measurements due to trajectory selection, and therefore, they presumably measure the total true intermittency.

\subsubsection{Flatness}
The flatness of the velocity structure functions has been computed for all cases. It is reported in figure \ref{fig_expoSp}(b). As already observed \cite{Mordant,LevNaso,Chevillard} it exhibits a plateau at small scales, followed by a rapid decrease around $\tau/\tau_k$. The plateau values provide the flatness of the time derivatives, that are clearly higher than Gaussian. The value of the plateau depends upon the Reynolds number and appear to decrease with increasing Reynolds number. This is at variance with what is observed in numerical simulations \cite{LevNaso} and we interpret this as a signature of the finite resolution of our measurements. Indeed, as previously discussed, our velocity fields are naturally filtered at a resolution fixed by the spatial resolution of our measurements. For very small $\tau$, the transition to regularity we observe is thus only governed by the cut-off time corresponding to such resolution, and we cannot access the true 
flatness, except for case A, where the spatial resolution is of the order of the Kolmogorov scale.  
\subsection{Lagrangian power structure functions}
\subsubsection{Time scaling}
\begin{figure}
(a)\includegraphics[width=0.47\textwidth]{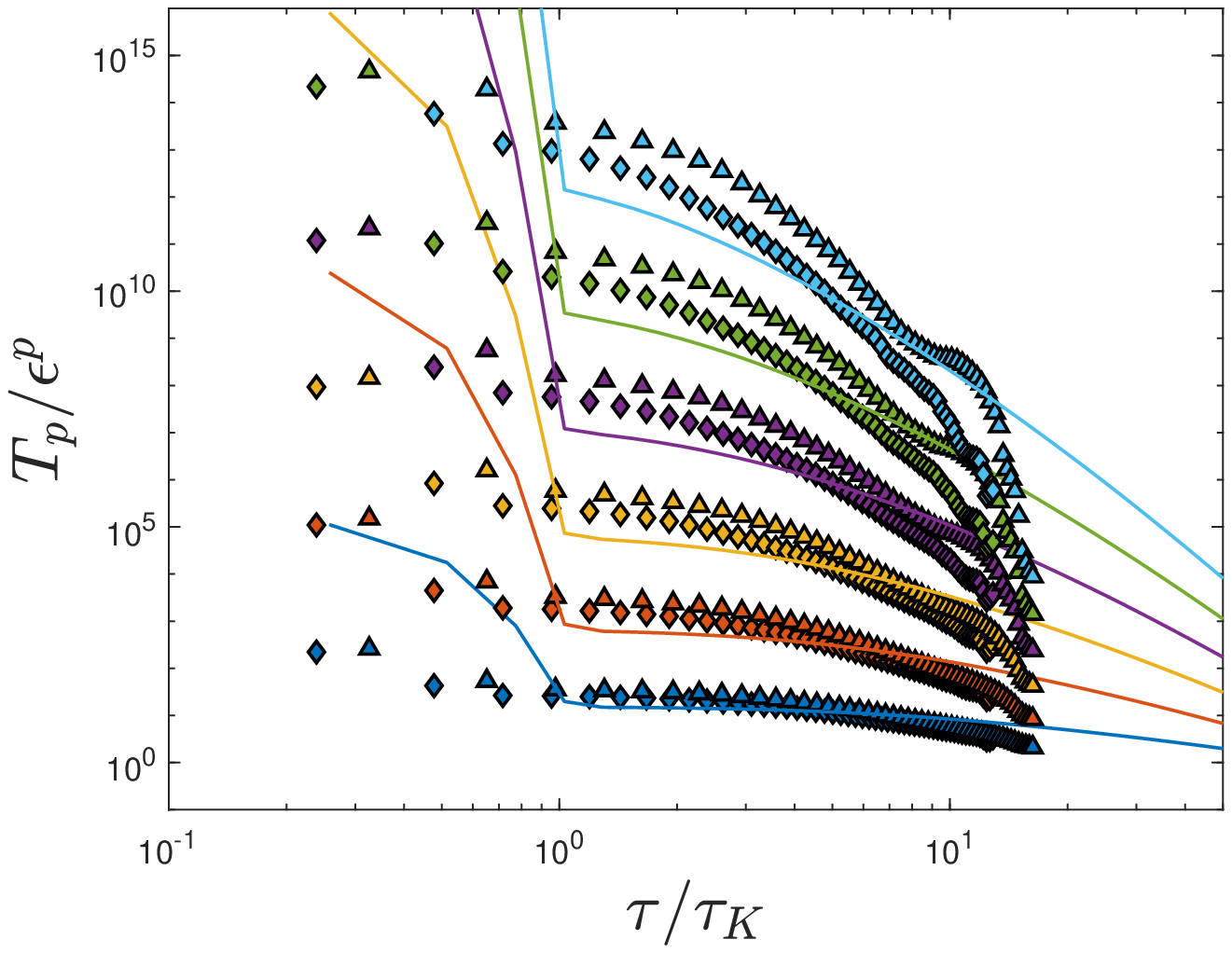}
(b)\includegraphics[width=0.47\textwidth]{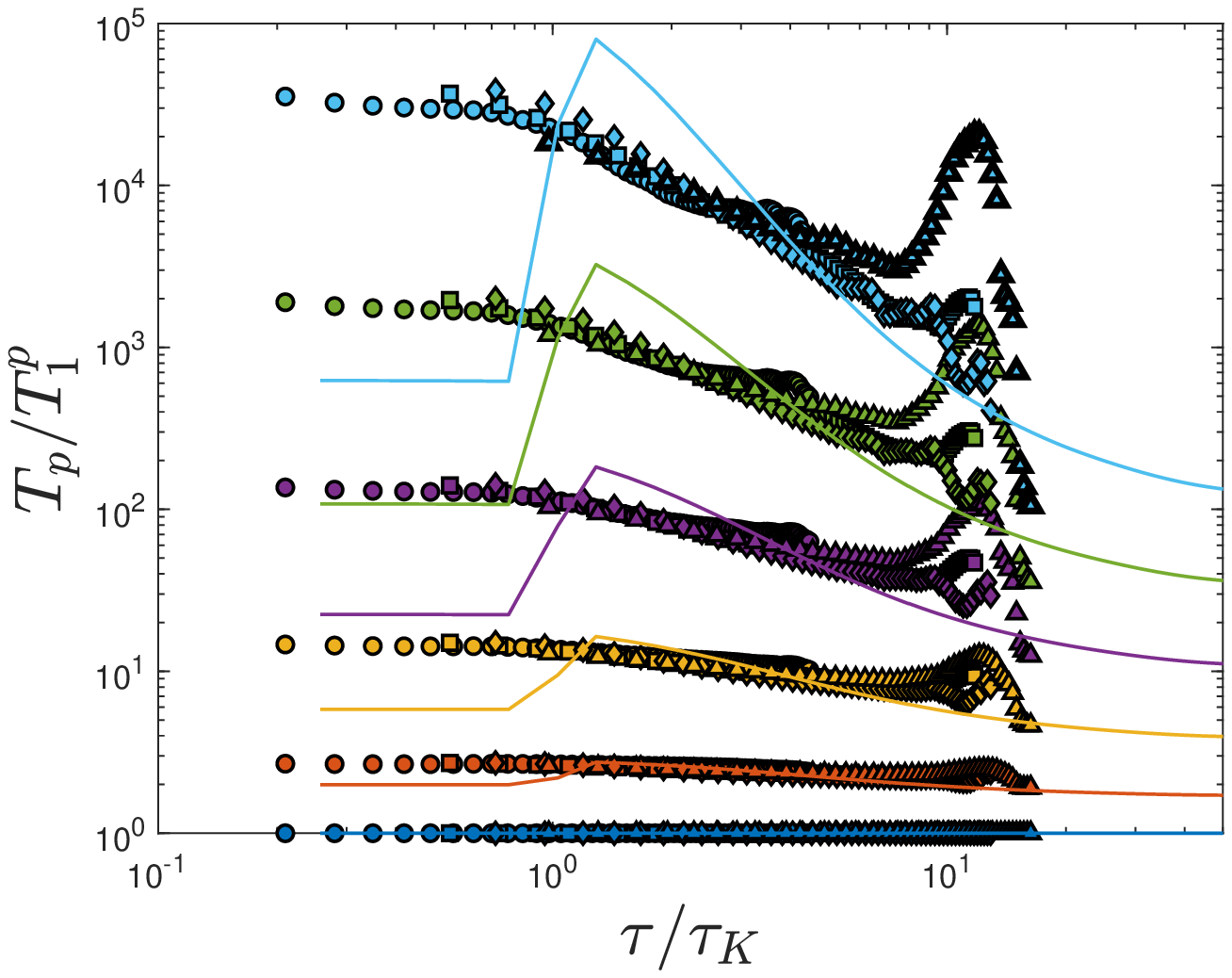}
\caption{Power structure functions ($T_p$) of order $p = [1 : 6]$ (blue, red, yellow, magenta, green, and cyan, respectively) as a function of $\tau/\tau_K$, where $\tau_K$ is the local Kolmogorov time. (a) $T_p$ scaled by $\epsilon^p$ for cases C (diamond), and D (triangle), where $\epsilon$ is the local energy dissipation rate; (b) $T_p$ scaled by $T_p(\tau = 1)$ for cases A (circle),  B (square), C (diamond), and D (triangle).\label{fig_Tp}}
\end{figure}

Figures \ref{fig_Tp}(a) reports the Lagrangian power structure functions defined in equation (\ref{Sp_P}) scaled  as a function of time over local Kolmogorov time. The curves are non-universal and do not collapse in these local variable, so that for clarity we only plot the structure functions corresponding to the highest experimental Reynolds number.
The Lagrangian power structure function are characterised by two regimes: 
\par as $\tau/\tau_K\to 0$, a steep increase towards a value that depends on the value of $R_\lambda$, and provides a filtered  estimate of the irreversibility indicators $R_p$.
\par for $\tau/\tau_K>1$, they display a milder decrease as $\tau$ increases. Such mild decrease is however significantly different from a plateau $T_p\sim \epsilon^p$, providing a clear indication of the intermittency of the Lagrangian power structure function. To try and detect a possible scaling regime in the inertial regime, we follow what was done in the Eulerian case for the local energy and dissipation structure functions \cite{Debue1} and computed the rescaled power structure functions $T_p/T_1^p$ as a function of time over local Kolmogorov time, shown in figure \ref{fig_Tp}(b) . Quite remarkably, we observe that like in the Eulerian case, all experimental cases collapse onto universal curves. Unlike in the Eulerian case, the universal curves do not coincide with the DNS curve, that vary more smoothly and decay faster. The universal curves display two regimes:
\par as $\tau/\tau_K\to 0$, a saturation towards a plateau.
\par as $\tau/\tau_K>1$, a transition towards a well defined scaling regime. Such scaling regime is not observed in the DNS case. We have computed the relative scaling exponents $\chi(p)-p\chi(1)$ of such scaling regimes and plot them in figure \ref{fig:compaMFR}(b). They are also reported in table  \ref{tablexpo}.

\subsubsection{Scaling exponents}
\begin{figure}
(a)\includegraphics[width=0.47\textwidth]{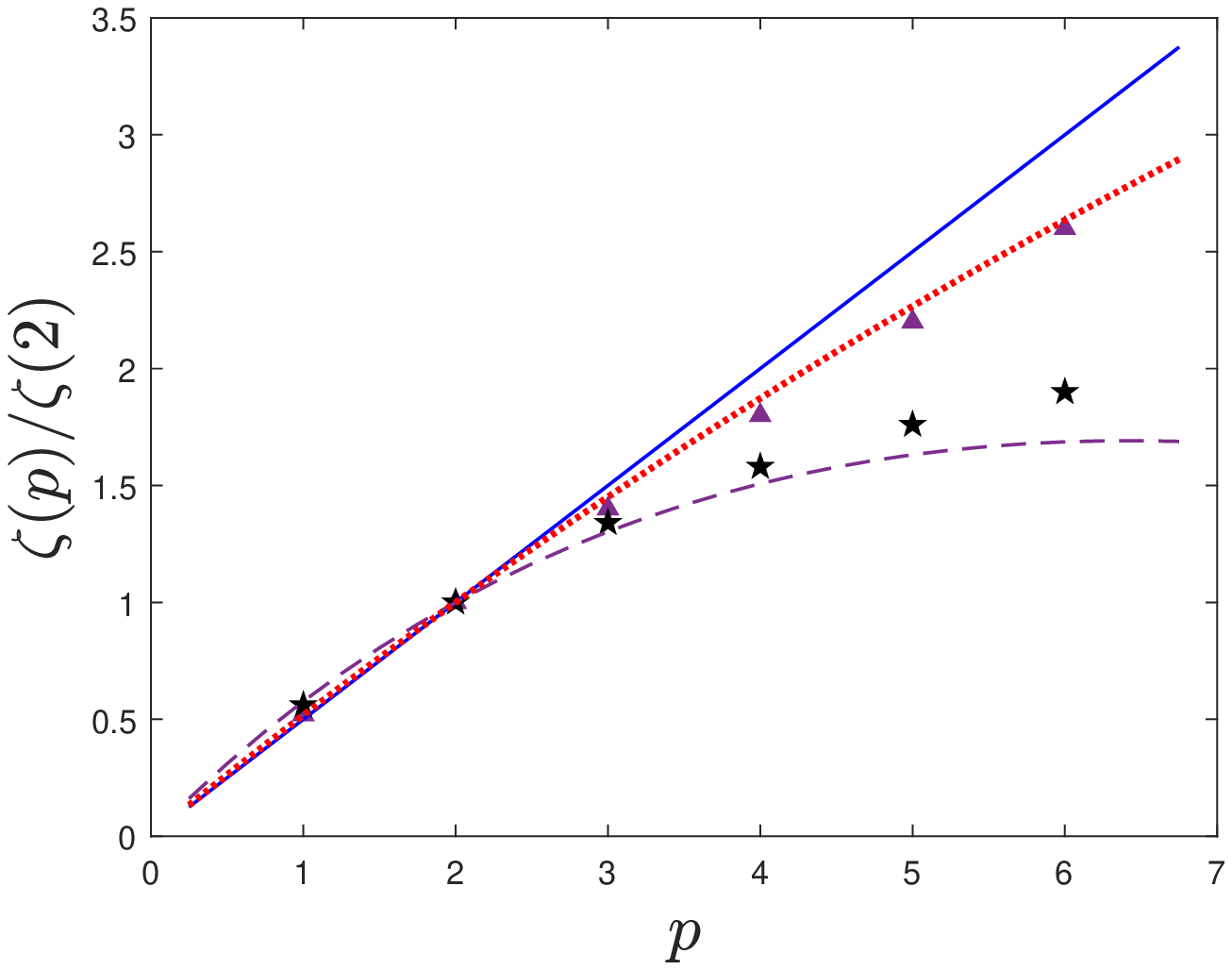}
(b)\includegraphics[width=0.47\textwidth]{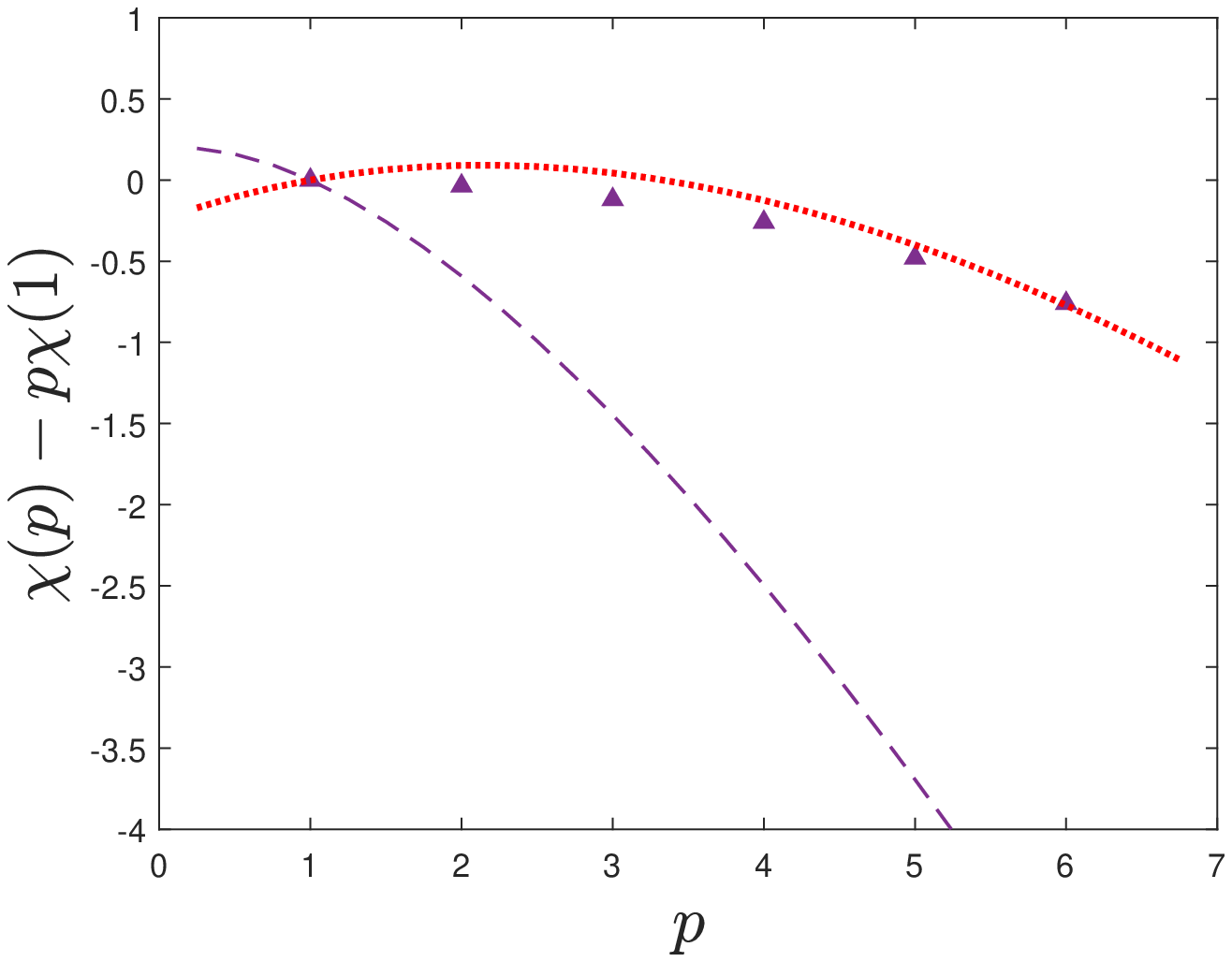}
\caption{Scaling exponents of order $p = [1 : 6]$, (a) for velocity structure functions $\zeta(p)/\zeta(2)$, (b) for power structure functions $\chi(p)-p\chi(1)$.
Magenta triangles: exponents obtained through ESS fits. Black stars: exponents measured by \cite{Mordant}  at $Re_{\lambda} = 510$ in another von K\'{a}rm\'{a}n set-up.   
The lines indicate the predictions from MFR log-normal models with  $C(h)=(h-a)^2/2b$:  blue line with $a=1/3$, $b=0$; red dotted line: $a=0.4$, $b=0.015$ magenta dashed line $a=0.35$, $b=0.045$.  }
\label{fig:compaMFR}
\end{figure}

To  understand their behaviour, we may use a straightforward adaptation  of the multifractal argument of Cencini et al. \cite{Biferale}, to find that the scaling exponents of the Lagrangian power scales like:
\begin{equation}
\chi(p)=\min_{h}\left(\frac{p(3h-1)+C(h)}{1-h}\right).
\label{MFRpower}
\end{equation}
This can be seen as the Lagrangian correspondence of the scaling exponent of the inertial dissipation, that are given by $\chi(p)^E=min_{p}\left(p(3h-1)+C(h)\right)$ \cite{Dubrulle}.
The prediction given by equation (\ref{MFRpower}) reported in figure \ref{fig:compaMFR}, for $C(h)$ given by log-normal models with the same values of the parameter than for the velocity structure function experiments. We see that the model coming for the Eulerian measurements does not fit well the exponents, but that the model with a lower intermittency parameter $b=0.015$ provides a good fit of the exponents. This is a confirmation that by sampling only long trajectories, we consider a portion of the flow with more regularity and less intermittency. Note that a similar intermittency reduction was observed in the Eulerian case, by conditioning statistics to areas of low local inter-scale energy transfers \cite{Dubrulle}.

\subsection{Irreversibility indicators and Reynolds dependence}
\begin{figure}
\centering
(a)\includegraphics[width=0.47\textwidth]{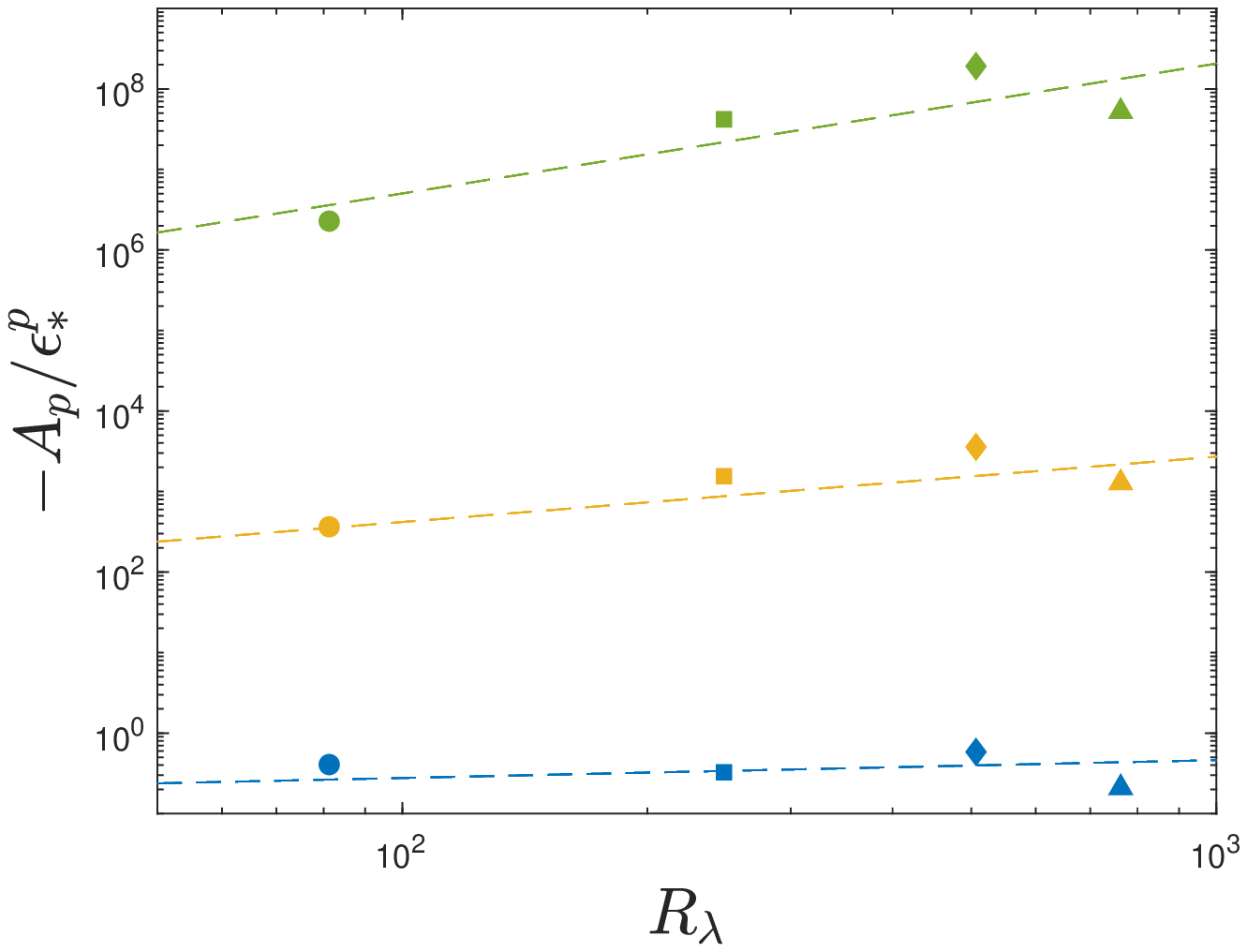}
(b)\includegraphics[width=0.47\textwidth]{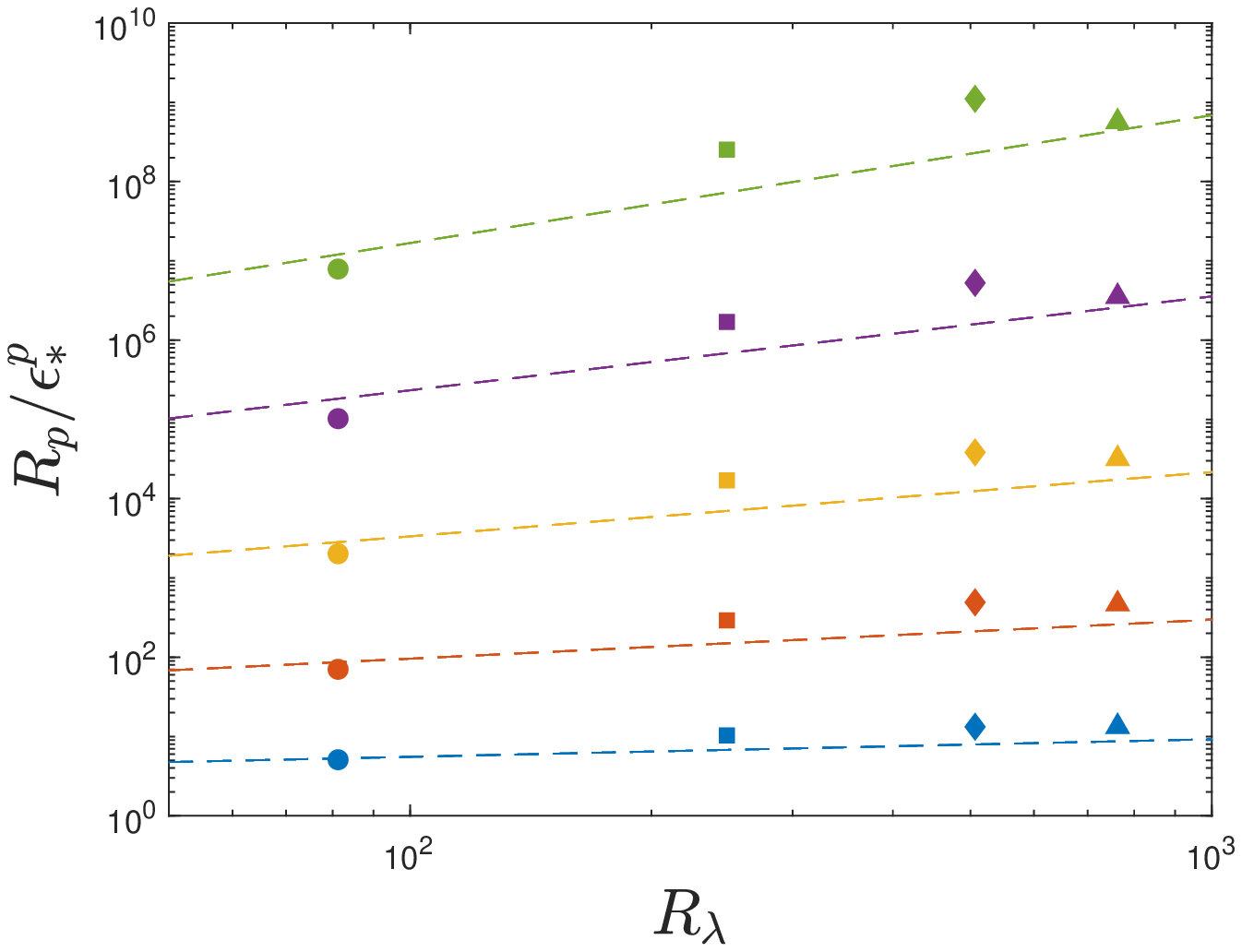}
\caption{
Irreversibility indicators (signed moments of the Lagrangian power $A_p$ and Lagrangian power moments $R_p$) for cases A (circle),  B (square), C (diamond), and D (triangle), versus the Taylor Reynolds number $R_\lambda$.
(a) $A_p$ of order $p = [1, 3, 5]$ (blue, yellow, and green, respectively); (b) $R_p$ of order $p = [1 : 5]$ (blue, red, yellow, magenta, and green, respectively).
The dashed lines are the power laws $R_\lambda^{1.4\alpha(p)}$, where $\alpha(p)$ is computed using the log-normal model $C(h)=(h-a)^2/2b$ with $a=0.4$ and $b=0.015$. \label{fig_Biferale}}
\end{figure}

To get the best estimate of the irreversibility indicators $R_p$ and $A_p$, we used Trackfitted data. Values are reported in figure \ref{fig_Biferale}(a) for the $A_p$ and in figure \ref{fig_Biferale}(b)  for the $R_p$. We see that all irreversibility indicators $A_p$ are negative.
For stationary measurements we should expect that $A_1=0$. In the present study we observe that $-A_1/\epsilon_*$ is of the order of a few tens of percent providing the degree of stationarity or convergency of our data. Given such numbers, we did not attempt to estimate very precisely the Reynolds number dependance of $R_p$ and $A_p$ \cite{Jucha,Biferale}.  Instead, we performed a consistency check of our data, by comparing the trends we observe with the predictions of the MFR theory which gives \cite{Biferale} 
\begin{eqnarray}
R_p&\sim& Re^{\alpha(p)},\nonumber\\
\alpha(p)&=&\min_{h}\left(\frac{p(2h-1)+C(h)}{1+h}\right),
\label{MFRIrreversibility}
\end{eqnarray}
where $C(h)$ is the same function that for the velocity increments. Taking into account the dependence of $R_\lambda$ with $Re$ in our data, we thus plot in figure \ref{fig_Biferale} the power laws $R_\lambda^{1.4\alpha(p)}$. We see that it provides a fair trend for our data. We also reported these exponents in table \ref{tablexpo}. They are smaller than what has been usually reported in the literature, meaning that the irreversibility is milder and grows much slower with increasing Reynolds number. This is an indication that irreversibility is associated with areas of larger irregularity, in agreement with current beliefs \cite{eyinkstoch}.\\

\begin{table}[h]
\caption{\label{tab_exp_small_tau} Summary of the exponents of the Lagrangian  structure functions  measured in experiments ($2^{nd}$ and $3^{rd}$ rows), or computed using a multifractal log-normal model 
with $C(h)=(h-a)^2/2b$,  $a=0.4$, $b=0.015$ ($5^{th}$, $6^{th}$, and $7^{th}$ rows).}
\begin{ruledtabular}
\begin{tabular}{|c|c|c|c|c|c|c|}
Measured Exponent &p=1&p=2&p=3&p=4&p=5&p=6\\
\hline
$\zeta(p)/\zeta(2)$ &0.5 &1 &1.4 &1.8 &2.2 &2.6 \\
\hline
$\chi(p)-p\chi(1)$ &0 &-0.04 &-0.12 &-0.26 &-0.48 &-0.76\\
\hline\hline
MFR prediction &p=1&p=2&p=3&p=4&p=5&p=6\\
\hline
$\zeta(p)/\zeta(2)$ &0.52 &1 &1.45 &1.87 &2.27 &2.63  \\
\hline
$\chi(p)-p\chi(1)$ &0 &0.09 &0.04 &-0.12 &-0.40 &-0.77\\
\hline
$1.4\alpha(p)$  &0.22&0.49 &0.81 &1.18 &1.61 &2.10\\
\end{tabular}
\end{ruledtabular}
\label{tablexpo}
\end{table}

\section{\label{conclusions}Conclusions}
4D-PTV measurements using very dense particle seeding and fast cameras are a relatively novel technique that allows in principle to get simultaneous Lagrangian and Eulerian measurements resolved both at the Kolmogorov time and scale, thereby opening new perspectives regarding experimental exploration of issues such as intermittency or irreversibility. In practice, such measurements are polluted by unavoidable experimental noise, both in time and space, and it is not clear how this will impede our capacity to improve our understanding of turbulence. In the present paper, we have explored such issues at different Reynolds number using two methods to deal with the temporal noise in our data. A first method uses a linear optimisation using a regularised B-spline algorithm. In contrast, a cruder noise removal using Gaussian filtering provides the same result than the velocity structure functions at large scale, but is not able to remove the small scale noise feature. This means that relatively elaborate noise reduction processes are necessary to be able to use the whole range of temporal scale in such kind of measurements.\\

Another difficulty was identified related to the length of our trajectories: due to our limited space measurement volume, we were able to track trajectories that do not exceed a few hundred of the inverse of the acquisition frequency. This means that we have limited access to the inertial range, if we want to keep measurements that are resolved in time. This issue could possibly be solved using a larger measurement volume or a velocity field with a mean velocity that is closer to zero. In our von K\'{a}rm\'{a}n flow, we have sought to achieve the smallest mean velocity by taking measurements at the center of the tank, but high variability of the large scale flow, due to organised coherent structures lying in the shear layer made it difficult to achieve this goal. This means that we should maybe try another geometry, devoided of shear layer.\\

On the other hand, we also found that trajectories of small extent (in our case smaller than about 50 $dt$) were polluted by a noise that we could not fully remove, even with our involved algorithm. This is probably due to the fact that such trajectories correspond to particle visiting areas of large irregularity, making it difficult for the particle tracking algorithm to follow them. By removing such small trajectories from our analysis, we thus introduced a bias towards more regular regions. Interestingly enough, we found that our measurements were then compatible with less intermittency and irreversibility.

Finally, we found that our measurements were all compatible with the MFR theory, and a log-normal multi-fractal spectrum. We further extended the analogy between the Lagrangian and Eulerian case by introducing a Lagrangian power structure function, that parallels the Eulerian structure function for inter-scale energy transfers defined in \cite{Debue1,Dubrulle}.\\

From a physical point of view, our most remarkable result seems the possibility to choose the degree of irregularity of the flow based on the length of the trajectories. It would then be interesting to  study the topology and properties of the corresponding Eulerian field, based upon a suitable interpolation onto a regular grid. We leave this for future work.

\begin{acknowledgments}
This work has been funded by the ANR, project EXPLOIT, grant agreement no. ANR-16-CE06-0006-01, and project ECOUTURB grant agreement no. ANR-16-CE30-0016.  Access to the HPC resources were provided by IDRIS (institut du d\'{e}veloppement et des ressources en informatique scientifique) under
the allocation A0062A01741  made by GENCI (Grand Equipement National de
Calcul Intensif).

\end{acknowledgments}

\bibliography{biblio}
\end{document}